\documentclass[conference,10pt]{IEEEtran}

\ifCLASSINFOpdf

\fi

\usepackage[subrefformat=parens,labelformat=parens,caption=false,font=footnotesize]{subfig}
\usepackage{array}
\usepackage{amsmath}
\usepackage{breqn}
\usepackage[nolist,nohyperlinks]{acronym}
\usepackage{graphicx,wrapfig,lipsum}
\usepackage{rotating}
\usepackage{balance}
\usepackage{amsmath, amssymb, amsthm}
\usepackage{stmaryrd}
\usepackage{tikz}
\usepackage{verbatim}
\usepackage{url}
\usepackage[utf8]{inputenc}
\usepackage[english]{babel}

\usepackage{multicol}
\usepackage{xr-hyper}
\newtheorem{rmk}{Remark}[]

\usepackage{scalerel}
\usepackage{multicol}
\newtheorem{definition}{Definition}
\usepackage[noadjust]{cite}
\usepackage[normalem]{ulem}
\usepackage{color}
\usepackage{dsfont}
\usepackage{bm}
\usepackage[short]{optidef}
\usepackage{diagbox}
\usepackage{enumitem}
\usepackage{slashbox}
\usepackage[lined,algonl,boxed]{algorithm2e}
\usepackage{multirow}
\usepackage[T1]{}
\usepackage{color, colortbl}
\usepackage{babel}
\usepackage{verbatim}
\usepackage{textcomp}
\usepackage{tabulary}
\usepackage{booktabs}
\usepackage{pgfplots}
\usepackage[export]{adjustbox}
\usepackage{textcomp}

\SetCommentSty{mycommfont}

\definecolor{Gray}{gray}{0.95}
\definecolor{LightCyan}{rgb}{0.8,0.85,1}
\definecolor{LightBlue}{rgb}{0.6,0.6,1}

\setlist{nosep}

\usepackage{mathtools}

\pgfplotsset{compat=1.18}

\begin{document}
\title{Data-driven Energy Efficiency Modelling in Large-scale Networks: An Expert Knowledge and ML-based Approach}

\author{
\IEEEauthorblockN{David L\'{o}pez-P\'{e}rez$^{a}$, 
Antonio De Domenico$^{b}$, 
Nicola Piovesan$^{b}$, and 
M\'{e}rouane Debbah$^{c}$}
\IEEEauthorblockA{$^{a}$\emph{Universitat Polit\`ecnica de Val\`encia, Spain}}
\IEEEauthorblockA{$^{b}$\emph{Huawei Technologies, Boulogne-Billancourt, France}}
\IEEEauthorblockA{$^{c}$\emph{Khalifa University of Science and Technology, Abu Dhabi}}
}

\maketitle

\begin{abstract}

The energy consumption of mobile networks poses a critical challenge. 
Mitigating this concern necessitates the deployment and optimization of network energy-saving solutions, 
such as carrier shutdown, 
to dynamically manage network resources. 
Traditional optimization approaches encounter complexity due to factors like the large number of cells, stochastic traffic, channel variations, and intricate trade-offs. 
This paper introduces the \ac{SRCON} framework, a novel, data-driven modeling paradigm that harnesses live network data and employs a blend of \ac{ML}- and expert-based models. 
These mix of models accurately characterizes the functioning of network components, 
and predicts network energy efficiency and \ac{UE} quality of service for any energy carrier shutdown configuration in a specific network.
Distinguishing itself from existing methods, 
\ac{SRCON} eliminates the reliance on expensive expert knowledge, drive testing, or incomplete maps for predicting network performance. 
This paper details the pipeline employed by \ac{SRCON} to decompose the large network energy efficiency modeling problem into \ac{ML}- and expert-based submodels. 
It demonstrates how, 
by embracing stochasticity, 
and carefully crafting the relationship between such submodels, 
the overall computational complexity can be reduced and prediction accuracy enhanced. 
Results derived from real network data underscore the paradigm shift introduced by \ac{SRCON}, 
showcasing significant gains over a state-of-the-art method used by a operator for network energy efficiency modeling.
The reliability of this local, data-driven modeling of the network proves to be a key asset for network energy-saving optimization. 

\end{abstract}

\section{Introduction}
\label{sec:intro}

The energy consumption of mobile networks is concerning.
Recent studies have shown that, 
while \ac{3GPP} \ac{NR} deployments are 4$\times$ more energy efficient than \ac{3GPP} \ac{LTE} ones,
they consume up to 3$\times$ more energy~\cite{Huawei2020}.
The increased energy consumption is primarily attributed to the deployment of denser networks handling new frequency bands with wider bandwidths through radios\footnote{
A radio denotes a radio unit. 
A wideband radio can manage multiple carriers/cells, 
and a \ac{BS} site may comprise one or multiple radios.} 
equipped with additional \ac{RF} chains to enhance beamforming and multiplexing capabilities~\cite{lopezperez2022survey}.
The 4$\times$ energy efficiency increase is far from the 100$\times$ target of the \ac{ITU} for \ac{5G} networks~\cite{M.2083-0}.
The 3$\times$ larger energy consumption is posing a threat on the environmental and business sustainability of cellular networks, 
where the energy cost of a mobile network already accounts for 23\,\% of the total operator cost~\cite{GSMA20205Genergy}.


Since the \ac{RAN} is the biggest contributor to energy consumption in a mobile network, 
with an average share of 73\,\%~\cite{GSMA20225Genergy},
the mobile industry has been working in solutions to enhance network energy efficiency. 
One key tool to minimize network energy consumption, 
while meeting the quality of experience of end-users, 
is to tailor network resources to \ac{UE} requirements~\cite{lopezperez2022survey}. 
To this end, 
the \ac{3GPP} has defined various network energy saving features to facilitate the implementation of different types of shutdown solutions, 
i.e. symbol, channel and carrier shutdown,
which allow the online (de)activation of time, space and frequency network resources, 
respectively~\cite{NGMN2023EnergyEfficiency}.


The potential of these network energy saving solutions is, however, suboptimally exploited in current \ac{4G} and \ac{5G} network deployments due to the complexity of their optimization.
The modelling and further optimization of network energy saving solutions are intricate problems, 
largely unsolved in both industry and academia.
The main challenges emanate from 
(i) the large number of cells and parameters per cell to configure, 
(ii) the stochastic and non-stationary nature of end-user traffic demands and the wireless channel, and
(iii) the intricate coupling/trade-offs between energy consumption and network/UE performance (in terms of coverage quality, throughput, reliability and latency)~\cite{NGMN2023EnergyEfficiency}.

{\color {black} 
Effectively optimizing network energy saving solutions, 
not only enhances network energy efficiency and reduces operator costs, 
but also significantly contributes to environmental sustainability.}

\subsection{Literature review}
\label{sec:intro:literature}

{\color {black} 
Various methodologies have been employed to tackle the issue of optimizing network energy saving solutions.
None of these techniques, however, have completely overcome the challenges at hand.
In the following, 
we review the most common approaches to network energy saving modelling and optimization, 
introducing the role of \ac{ML}.}

\subsubsection{Expert knowledge}

Currently, 
expert knowledge still remains a primary means for optimizing network parameters.
Experts rely on cell-level statistics~\cite{3GPPTS32.450}, 
gathered from \acp{BS}, 
as well as \acp{CDR}~\cite{7008502} and \acp{DT}~\cite{Unlaut2022} measurements to evaluate network performance and \ac{UE} quality of service at various locations,
and take decisions. 
However, \acp{CDR} and \acp{DT} necessitate large measurement campaigns and specialized equipment, 
which make them economically expensive and diminish their worth. 

To alleviate this cost challenge,
the \ac{3GPP} introduced minimization of \ac{DT}~\cite{6211483},
which employs \ac{UE} geolocated radio measurements to inexpensively assess network performance.
However, only a few \acp{UE} enable this feature nowadays due to privacy concerns, 
and they usually do it for a short period of time. 

Nevertheless, 
it is essential to acknowledge that cell-level statistics, \acp{CDR}, \acp{DT}, and minimization of \ac{DT} alone are not sufficient for network optimization. 
These metrics assess network performance only for the specific configuration ---or, at most, a few configurations--- set during the measurement campaigns.
They do not provide insights into network performance across the billions of other potential network configurations.
Consequently, 
state of the art trial and error optimization approaches based on expert knowledge as well as cell-level statistics, \acp{CDR} and \ac{DT}-based data cannot accommodate for the requirements of today’s real world 4G/5G large-scale network optimization. 

\subsubsection{System-level simulation}

System-level simulation is another widely used option for assisting network optimization~\cite{Huawei2018}.
Precise modeling of the propagation environment and protocol stack are crucial to obtain an accurate evaluation and generalization of network performance to any network configuration. 

Available tools for modeling the wireless propagation environment range from basic models that solely rely on statistics to advanced ray-tracing systems~\cite{8438326}. 
However, these precise tools often require hard-to-obtain map information and time-consuming operations, 
making them expensive.
Examples of such radio propagation prediction tools are~\cite{Forsk_Atoll,Ranplan_Wireless}. 

With regard to the protocol stack,
frameworks such as NS-3 5G-LENA~\cite{KOUTLIA2022} and its evolution~\cite{Mezzavilla2018} have gained popularity due to their capabilities. 
Nevertheless, they fail at modelling the complexity of true products. 
For example, 
they do not take into account the heterogeneity ---and different performance--- of off-the-shelf \ac{BS} and \ac{UE} products. 

\subsubsection{Theoretical tools}

Theoretical tools have also been developed to aid network energy efficiency optimization. 
Importantly, 
optimal shutdown policies for a single server with bursty traffic were derived in \cite{Bingjie2017}. 
The optimal shutdown policy was shown to be a two-threshold policy, 
with one threshold to drive the sleep procedure and a different one to manage the cell wake up.
Even if there is no theoretical results available for more than one server, 
the findings of this work are in line with today´s carrier shutdown solutions, 
which implement two different sets of conditions for shutdown and reactivation (see Section \ref{sec:sd_logic}).

When theoretically studying more complex multi-cell macrocellular networks, 
we should highlight the coverage, capacity and energy efficiency trade-offs derived in \cite{Bjornson2015optimal}, \cite{Yu2015} and \cite{LopezPerez2021} for large-scale cellular networks with \ac{mMIMO}, \ac{CA} and joint \ac{mMIMO} and \ac{CA} capabilities, respectively.

In \cite{Soh2013},
the authors compared the energy efficiency trade-offs associated with a macrocellular network equipped with shutdown capabilities to those of a small cell network, 
discussing the pros and cons of these two deployment strategies
The further analysis in~\cite{Celebi2019} showed the benefits of deploying advanced sleep modes at the small cells, 
proposing multiple carrier shutdown strategies,
and comparing them in terms of complexity, blocking rate probability, throughput and energy efficiency.

From an optimization perspective,
the authors of~\cite{Feng2017TWC} investigated how to fine-tune network energy efficiency by jointly managing the long-term cell activation, \ac{UE} association and power control in a heterogeneous network with \ac{mMIMO} capabilities, 
and proposed a distributed solution based on game theory, 
proven to converge to the Nash equilibrium.

The work in~\cite{Chiaraviglio2017} also modelled hardware failure rate due to cell (de)activation, 
and proposed a heuristic to control the \ac{BS} shutdown, 
which minimises the acceleration over time of the hardware failure rate, 
while satisfying the \ac{UE} quality of service demands.

Although the models produced and insights gained from these studies are valuable, 
their reliance on numerous assumptions ---stemming from the theoretical nature and complexity of large-scale networks---limits their applicability. 
For instance, the carrier shutdown scenarios explored in this research significantly differ from those implemented in actual network environments. 
Therefore, these theoretical approaches fall short when applied to the optimization of energy saving solutions in real-world network settings.

\subsubsection{Data-driven modelling and optimization}

{\color {black} 
Given the limitations noted, 
there is an increasing interest in leveraging \ac{ML} to develop more scalable and flexible approaches for network optimization~\cite{Dai2019}. 
Recognizing this potential, 
the \ac{3GPP} has initiated a shift towards a new paradigm in data-driven network modeling and optimization. This effort focuses on identifying the crucial measurements and protocols necessary to devise advanced data-driven strategies for localized network optimization~\cite{3GPPTR37.817}. 
By applying \ac{ML} techniques, 
the aim is to predict network performance accurately and improve various dimensions of network functionality, notably energy efficiency~\cite{3GPPTR37.816}.}

{\color {black} 
To date, 
such explorations into data-driven network modeling and optimization have predominantly focused on developing \ac{ML} models that harness data from numerical simulations and test environments.
These models are not just academic exercises,
but are intended to introduce new mechanisms at the physical layer of communication systems~\cite{Zhang2019}, 
as well as to refine the functionalities at the network's upper layers~\cite{Luong2019}.
Significant advancements in physical layer mechanisms include the design of neural receivers specifically for 5G multi-\ac{UE} \ac{MIMO} configurations~\cite{10008602}, 
alongside the introduction of deep learning-based synchronization~\cite{10001353} and decoding schemes~\cite{10008601}. 
Intelligent \ac{MIMO} channel estimators have also emerged as a notable example application. 
These \ac{ML} estimators are designed to mitigate the complexity inherent in channel detection procedures while minimizing performance degradation~\cite{9852760}.
In terms of radio resource management functionalities, 
new deep learning approaches aim at streamlining wireless resource allocation algorithms, 
showcasing a significant leap in managing the complexities of \ac{MAC} layer network operations~\cite{9737018}.}

{\color {black} 
In contrast, 
only a few studies have explored the use of real-world network data for modeling and enhancing large-scale networks, 
the subject at hand. 

\paragraph{Traffic prediction}
Within this domain, 
substantial strides have been made in data-driven traffic prediction with real-world data employing various \ac{ML} techniques, 
such as \acp{LSTM}, \acp{CNN}, and \acp{GCN}~\cite{Feng2018,zhang2019deep,Wu2020connecting,piovesan2021}. 
These methodologies are pivotal for facilitating proactive network optimization efforts. 
Cutting-edge traffic prediction approaches are increasingly leveraging \acp{GAN}, 
combined with traffic data and expert knowledge~\cite{2023ZhangTraffic, 2023HuiTraffic}. 
Integration of urban environment knowledge graphs have further enriched traffic prediction models, 
improving their accuracy and thus the quality of network optimization.

\paragraph{Anomaly detection}
The application of data-driven modeling using real-network data has also been significantly explored for anomaly detection and network alarm predictions.
Advanced \ac{ML} models, 
similar to those used for traffic prediction, 
have been adapted to detect anomalies by learning from vast datasets of network activity. 
These models are trained to recognize patterns indicative of potential issues, 
enabling network operators to preemptively address problems before they escalate~\cite{7149014,9124782,9143579,10410176}.

\paragraph{Cell rate and \ac{UE} spectral efficiency predictions}
Gijon et al. assessed cell throughput prediction accuracy using real network \ac{KPI} counters, 
exploring a variety of \ac{ML} methodologies such as support vector regression, k-nearest neighbors, and decision trees and \ac{ANN}-based ensemble approaches~\cite{9896982}. 
Advancing this research, 
Xing et al. introduced a neural network framework aimed at modeling the interference and spectral efficiency of \ac{DL} \ac{5G} \ac{mMIMO} transmissions, 
employing \ac{RSRP} data from \acp{DT}~\cite{Zheng2022}. 
Their work underscores the potential of \ac{ML} in accurately forecasting cell throughput and enhancing understanding of \ac{UE} spectral efficiencies.
However, these models lack the capability to dynamically forecast \acp{KPI} in response to changes in network configurations, 
revealing a significant gap in their adaptability to real-world optimization challenges. 
Furthermore, they face challenges in accurately predicting \ac{UE} rates, 
mainly because they do not adequately consider crucial factors such as bandwidth allocation per \ac{UE}, which are profoundly influenced by network energy-saving strategies.
 
\paragraph{Energy consumption and carbon emissions estimations}
To evaluate the energy-related performance of the \ac{RAN}, the mobile industry has defined several measurement methods and metrics for different network levels (network \cite{etsiES203228}, site \cite{ituL13502016}, \ac{BS} \cite{etsiTS103786}, and UE \cite{3gpp38840}) as well for different scenarios (dense urban, urban, and rural coverage) \cite{etsiES203228}, and services (enhanced mobile broadband, ultra-reliable low-latency communications, and massive machine type communications) \cite{3gpp28813}. More recently, Rappaport et al. have introduced the Waste Factor \cite{rappaport2024waste}, a new metric for quantifying energy efficiency in a wide range of circuits and systems applications, including data centers and RANs.

Li et al.'s study, 
utilizing network data from Nanchang, 
modeled carbon emissions from mobile networks and expanded these insights across China~\cite{2023LiNat}. 
The research used energy consumption and network traffic data from Nanchang, 
alongside \ac{BS} and mobile user counts across Chinese provinces, 
using simple models for extrapolation. 
Additionally, 
they introduced a \ac{RL} framework designed to shut down network cells as a means to prevent capacity over-provisioning. 
A significant limitation of the model, 
however, 
is its simplistic approach to energy-saving schemes ---cells are shut down as deemed necessary and \ac{UE} handovers are ideal--- coupled with an inability to evaluate \ac{UE} performance following changes. 
Maggi et al. have made a substantial contribution by introducing a sector-specific, closed-loop Bayesian method to optimize carrier shutdown thresholds, 
effectively reducing power consumption while maintaining network \acp{KPI}, 
particularly \ac{UE} rates~\cite{10188338}. 
This data-driven approach, 
validated in a live 4G network, 
emphasizes the practical application and effectiveness of their model. 
However, the focus on optimizing carrier shutdown thresholds, excluding the refinement of handover algorithms, and the model's limitation in simulating network behavior offline might present challenges in more complex scenarios where Bayesian optimization may fail to converge~\cite{2020Moriconi}.

\bigskip 

Drawing from the literature, 
it becomes evident that while significant strides have been made in network optimization through various methodologies, 
a crucial gap remains: a cohesive, data-driven modeling framework. 
Existing research, valuable though it is,often compartmentalize aspects of network performance, 
neglecting the intricate balance required between energy efficiency and maintaining optimal network \acp{KPI} in dynamic real-world scenarios. 
This highlights the urgent need for a comprehensive framework that not only bridges these divides, 
but also utilizes real-world data to thoroughly model and optimize the trade-offs between energy consumption and key network performance metrics. 
This includes considering the complex interplay between energy-saving strategies, mobility management, and other network dynamics. 
Such a framework promises to advance the current state of network optimization by providing a more accurate, adaptable, and practical approach to energy-saving solutions, 
thereby addressing the limitations identified in the existing body of work.}

{\color {black} 
\subsection{Research questions and objectives}

In light of the current state of the art, 
this paper sets forth a series of research questions and scientific objectives to delineate the research problem at hand, and move towards the creation of such comprehensive end-to-end framework:

\textbf{Research Questions:}
\begin{itemize}
\item
Is it feasible for a data-driven model, 
reliant solely on off-the-shelf network data ---no \acp{DT}--- to deliver accurate predictions of network energy consumption and user throughput? 
\item
How can we seamlessly blend machine learning algorithms with domain expertise to enhance energy efficiency modeling in large-scale networks? 
\item
What advancements in network energy efficiency prediction can this type of modeling framework introduce, and how does it measure up against existing methodologies? 
\end{itemize}

\textbf{Scientific objectives:}
\begin{itemize}
\item
Establish a Data-Driven Modeling Framework: 
Our goal is to devise a holistic framework that can precisely forecast network energy efficiency and user throughput using only inexpensive available network data. 
This requires the harmonization of machine learning (ML) technologies with expert insights to depict the multifaceted nature of large-scale networks.
\item
Consider Practical Network Energy Saving Solutions: 
The framework aims to furnish practical energy saving solutions into the large-scale network modelling for elevating energy efficiency in telecom networks.
\item
Validate the Framework with Real-World Data: 
We intend to prove the framework's utility by applying it to actual network scenarios, showcasing its capability for precise energy consumption predictions and its value in fostering user throughput when fine tuning energy saving solutions.
\end{itemize}
}

\subsection{Paper's contributions}

Building upon the research questions and scientific objectives outlined previously, 
this paper introduces a pioneering solution to the challenges of optimizing network energy-saving solutions. 
We present a novel data-driven framework for modeling {\color {black} network energy consumption and \ac{UE} throughput},  
referred to as \ac{SRCON}~\cite{SRCON2023Intro},
which uses measurement data from live networks to jointly fit 
(i) \ac{ML}-based, black-box and 
(ii) expert-based, white-box models\footnote{
In contrast to a black-box model,
in a white-box model logic, functioning and programming steps are transparent and known. 
As a result, it's decision making process is interpretable. 
} 
that can
\begin{itemize}
    \item 
    accurately characterize the functioning of the most relevant network components, and 
    \item
    generalise the modelling of {\color {black} network energy consumption and \ac{UE} throughput} to any possible network energy saving solution configuration,
    even those not observed in the training data.
\end{itemize} 

{\color {black} 
\ac{SRCON}, 
akin to digital twins, 
functions as a virtual representation of the physical system; in this instance, the network, 
acting as its digital counterpart. 
Digital twins are increasingly viewed as a vital paradigm for monitoring, controlling, and optimizing communication systems~\cite{9711524,9854866}. 
They offer a sandbox for testing new features, 
e.g.\ac{AI} solutions, 
potentially reducing the need for field data collection and algorithm testing.
Most digital twins often rely on \ac{IoT} sensors to proactively analyze system performance in real-time and make decisions accordingly~\cite{9429703}. 
A key challenge in digital twin is ensuring that virtual control optimization is safe and reliable, 
preventing incorrect decisions due to "model exploitation."~\cite{10234596} 
\ac{SRCON} diverges from typical digital twin frameworks ---and attempts to enhance reliability--- by not attempting to replicate network behaviors on a small timescale, 
e.g., milliseconds ---a task that is fundamentally impractical due to the significant modeling challenges faced in real-world large-scale networks. 
Unlike the relatively straightforward sensor data utilized in many digital twin applications, 
the network's 'sensors' are the \acp{UE} themselves, 
whose interactions and performance data are inherently multi-dimensional and complex to analyze. 
Given this intricacy, 
\ac{SRCON} instead is set to achieve a \emph{statistically indistinguishable} emulation of network behaviors. 
We should also underscore that,
unlike many digital twins focused on real-time optimization, 
\ac{SRCON}'s current goal is not real-time modeling and optimization but rather to establish foundational functions for such future optimization capabilities.
Further details on these aspects are provided throughout the document.}

{\color {black} It is important to acknowledge that our current scope focuses on network energy consumption and \ac{UE} throughput, pivotal \acp{KPI} for evaluating the efficiency of energy-saving solutions. 
However, we recognize the significance of other metrics, 
such as delay, 
Diving into the modeling of such metrics represents a promising avenue for future research, 
requiring additional modelling. 
We leave this to future work.} 

{\color {black} 
With respect to the available literature earlier reviewed in Section \ref{sec:intro:literature}, 
the novelty of this paper resides in the following aspects: 
\begin{itemize}
\item {Novel Modeling Approach}: Unlike existing literature, this work uniquely focuses on modeling network energy consumption and \ac{UE} throughput within the context of a practical carrier shutdown solution and its associated handover procedures across a vast, heterogeneous network of hundreds of 4G and 5G cells. This eliminates the need for expensive drive testing or ray tracing-based simulation tools.

\item {Decomposed Modeling Framework}: We introduce a novel methodology by segmenting the overall modeling challenge into distinct \ac{ML}- and expert-based subproblems. Outputs from expert-based models inform \ac{ML}-based models, enabling precise generalization of network energy consumption and \ac{UE} throughput predictions across any carrier shutdown configuration, including unencountered scenarios.

\item {Data and Model Construction Pipeline}: This paper provides a detailed account of the necessary data and the pipeline used to construct \ac{ML}-based models for network energy consumption and \ac{UE} throughput. This method is crucial for accurately capturing the varied performance characteristics of network products and end-user devices --- an approach not seen in current literature. Each problem necessitates a unique framework.

\item {Minimal Assumption Requirement}: A distinguishing feature of our modeling approach is its minimal dependence on the statistical assumptions often mandated by methods like Bayesian optimization, which typically assume Gaussian behavior. Our models' capability to provide reliable predictions without such assumptions significantly increases their robustness and adaptability across various network scenarios.

\item {Innovative Expert-Based ABM Model}: Introducing an innovative \ac{ABM} marks a significant advancement. This expert model thoroughly simulates network stochasticity and the dynamics of carrier shutdown across wide-area networks with manageable complexity. It includes handover procedures and approximates \ac{UE} transfers. The ABM enables precise predictions using established telecom theories, supplying inputs to \ac{ML} models for any configuration of carrier shutdown solutions and handover parameters.

\item {Data-Driven Comparative Results}: We conclude with empirical findings that underscore the extensive benefits of our modeling approach. A comparative evaluation against a benchmark tool, utilized by a leading network operator for carrier shutdown optimization, illustrates our models' superior performance and effectiveness.

\end{itemize}}

\bigskip 

The rest of this paper is structured as follows. 
In Section~\ref{sec:carriershutdown}, 
we introduce the basics of a state of the art carrier shutdown solution, 
with the corresponding formal definitions. 
In Section~\ref{sec:modellingChallengesAndSRCON}, we highlight the main challenges to model carrier shutdown performance in a large-scale network, 
and present the main logic behind our proposed modelling approach,
with a mix of black- and white-box modelling. 
In Section~\ref{sec:data}, we detail the different types of data,
which are available to us to carry on our modelling exercise. 
In Section~\ref{sec:pc_model} and~\ref{sec:rate_model},
we present the energy consumption and \ac{UE} rate models developed to investigate the energy efficiency of practical networks.
These models are constrained by ---tailored to---  the available data. 
In Section~\ref{sec:abm_model},
we present the main contribution of this paper, 
our expert-based, white-box, \ac{ABM},
which mimics the stochastic behaviour in a large-scale network with carrier shutdown,
and allows to generalise the network energy consumption and \ac{UE} rate predictions to any carrier shutdown and handover parameter configuration.
The outputs of this model are inputs to the models presented in Section~\ref{sec:pc_model} and Section~\ref{sec:rate_model}.
To demonstrate the generalization capabilities of our framework, in Section~\ref{sec:results},
we describe the scenarios and conditions under which our modelling approach has been tested in a real network,
and discuss the performance results. 
Finally, in Section~\ref{sec:conclusions}, we drawn the conclusions,
and highlight new research directions. 
\section{Carrier shutdown and modelling challenges}
\label{sec:carriershutdown}


Carrier shutdown allows to fully deactivate the cells/carriers\footnote{
Note the while cell refers to the area covered by a carrier operating at a given frequency with a given bandwidth, 
these two terms are used interchangeably in this paper.} 
mounted on a given radio unit by switching off most of its \ac{RF} and digital front-end components.
The base band unit and the interface for waking up the shutdown carriers remain active.
For more details on carrier shutdown, 
please refer to~\cite{NGMN2023EnergyEfficiency}.


Importantly, 
to enable a dynamic carrier shutdown operation,
the \ac{3GPP} has specified a number of related features~\cite{3GPPTS38.300}, 
building on the concepts of capacity booster cell and coverage cell,\footnote{
To ease the complexity of carrier shutdown, 
in a multi-layer network, 
cells are usually divided into two groups: 
Capacity booster cells are deployed for capacity enhancements, 
typically use a higher carrier frequency, 
and can be shut down. 
Coverage cells, instead, are deployed to provide blanket coverage, 
usually use a lower carrier frequency, 
and cannot be shut down.} 
among which we should highlight:
\begin{itemize}
    \item Cell pairing:
    Capacity booster cells and coverage cells can be paired for energy efficiency purposes, 
    where an  \ac{LTE} or an  \ac{NR} cell can be the coverage cell of a capacity booster cell.
    \item Autonomous shutdown:
    A capacity booster cell can take autonomous carrier shutdown decisions based on estimations,
    not only of its own number of connected \acp{UE} and \ac{DL} and \ac{UL} cell loads, 
    but also on available load information from paired neighbouring coverage cells exchanged through the respective X2/Xn interfaces.
    \item Reactivation:
    The paired coverage cell owning a capacity booster cell can autonomously request an inter-\ac{BS} cell reactivation over the X2/Xn interface based on its own load information.
    \item Shutdown information sharing:
    A coverage cell can inform all its neighboring cells of the (de)activation of one of its capacity booster cells.
\end{itemize}
The \ac{3GPP} efforts in this front still continue today~\cite{3GPPTR37.816,3GPPTR38.864}. 


With respect to the general functioning of carrier shutdown solutions,
it should be noted that carrier shutdown and wake up decisions are generally assessed by capacity booster cells and coverage cells, respectively, with a frequency in the order of seconds,
while the carrier shutdown duration may well vary from tens of seconds to minutes or even hours. 
The time that it takes to shutdown and wake up a carrier is around 3 seconds. 


Since carrier shutdown allows deactivating the entire cell, 
it enables deeper sleeps than symbol and channel shutdown, 
and in turn, larger energy savings. 
However, coverage and capacity losses can be significant, 
if carriers are not (de)activated in a coordinated manner across the network. 
For example, 
coverage holes may appear when shutting down a cell in those cases where no other cell can provide coverage to the former \acp{UE} of the shutdown cell. 
Capacity may also be affected, 
as a shutdown cell does not allow for the spatial reuse of spectrum, 
and thus the service quality of the \acp{UE} of the cells receiving the \acp{UE} of the shutdown cell may be compromised due to resource sharing. 
Similarly, 
the load in the cell receiving the \acp{UE} of the shutdown cell may grow after some time, 
and if the shutdown cell is not reactivated on time, they may also suffer from service quality degradation. 
This gives rise to an intricate trade-off between network energy consumption and network/\ac{UE} performance.


The target of this paper is to present a novel, data-driven modelling approach to assess this trade-off,
which can be later used to optimize carrier shutdown operations in practical networks. 
In the following, 
we detail the specific carrier shutdown logic assumed in this paper,
which is in line with that used in real solutions~\cite{NGMN2023EnergyEfficiency}. 

\subsection{Cell pairing}

As indicated earlier,
for energy efficiency purposes,
cells are divided between (higher carrier frequency) capacity and (lower carrier frequency) coverage cells,
and every capacity cell is paired with at least one coverage cell,
which should (i) provide service to the \acp{UE} of the capacity cell, 
should this one shut down,
and (ii) assist its wake up process. 
Thus, the paring process between capacity and coverage cells is capital to the overall carrier shutdown performance. 

The coverage cell paired with a capacity cell should be the coverage cell that has the largest coverage overlap with the capacity cell. 
Given the difficulty to predict the overlaps among cells in a planning stage, 
these relationships are usually computed based on inter-frequency measurements periodically reported by the \acp{UE}. 
In a nutshell, 
the following conditions should be met to consider a given inter-frequency cell as a pairing candidate:
\begin{itemize}
    \item
    The number of times that such coverage cell is reported by the \acp{UE} of the capacity cell is larger than a threshold, 
    and 
    \item
    the ratio of the number of times that such coverage cell is reported to the total number of measurement reports is larger than another threshold, 
    and
    \item
    among those measurement reports in which such cell is reported, 
    the fraction of reports with an \ac{RSRP} larger than an RSRP coverage threshold is larger than another threshold. 
\end{itemize}  
Although of importance,
the modelling of this capacity and coverage cell pairing process is out of the scope of this paper,
and it will considered as optimised and given in the following. 

\subsection{Reference carrier shutdown (de)activation logic}
\label{sec:sd_logic}

In the reference implementation used in this paper, 
assuming that a capacity cell, $c$, is paired with only one coverage cell, $b(c)$, 
{\color {black} and to maintain the \ac{QoS} of \acp{UE} affected by the transfer,}
carrier shutdown can only be autonomously activated in capacity cell, $c$, when 
\begin{itemize}
    \item 
    the number, $U^{\rm UE}_{c}$, of \ac{RRC} connected \acp{UE} in the capacity cell, $c$, is smaller than an entry threshold, $\chi^{\rm UE}_{c}$, and
    \item 
    the sum, $\Delta^{\rm DL}_{c,b(c)} = \Delta^{\rm DL}_{c} + \Delta^{\rm DL}_{b(c)} $, of used \ac{DL} \acp{PRB} in the capacity cell, $c$, and the paired coverage cell, $b(c)$, is smaller than another entry threshold, $\chi^{\rm DL}_{c,b(c)}$, and 
    \item 
    the sum, $\Delta^{\rm UL}_{c,b(c)} = \Delta^{\rm UL}_{c} + \Delta^{\rm UL}_{b(c)}$, of used \ac{UL} \acp{PRB} in the capacity cell, $c$, and the paired coverage cell, $b(c)$, is smaller than another entry threshold, $\chi^{\rm UL}_{c,b(c)}$.
\end{itemize}

In contrast, 
a coverage cell, $b(c)$, may wake up its paired capacity cell, $c$, in carrier shutdown when 
\begin{itemize}
    \item 
    the number, $U^{\rm UE}_{b(c)}$, of \ac{RRC} connected \acp{UE} in the paired coverage cell, $b(c)$, is larger than a leaving threshold, $\Psi^{\rm UE}_{c,b(c)}$, or
    \item 
    the number, $\Delta^{\rm DL}_{b(c)}$, of \ac{DL} \acp{PRB} in the paired coverage cell, $b(c)$, is larger than another leaving threshold, $\Psi^{\rm DL}_{c,b(c)}$, or 
    \item 
    the number, $\Delta^{\rm UL}_{b(c)}$, of used \ac{UL} \acp{PRB} in the paired coverage cell, $b(c)$, is larger than another leaving threshold, $\Psi^{\rm UL}_{c,b(c)}$.
\end{itemize}  
Time windows are used to average and smooth these statistics. 
As one can imagine,
the optimization of these thresholds plays a major role on energy savings, 
and are thus part of our model.  

{\color {black} 
The careful consideration of these algorithms ensures that the paired coverage cells can seamlessly support the communications of all \acp{UE} in the capacity cell that is shutting down, in addition to those of its already connected \acp{UE}, thereby maintaining \ac{QoS} without disruption. 
A bias could be added to the corresponding thresholds to ensure that a larger number of \acp{PRB} are reserved for the \acp{UE} being transferred. By ensuring that the paired coverage cells can accommodate the \acp{PRB} of the transferred \acp{UE}, we uphold the \ac{QoS}.}

\subsection{Reference UE transfer logic}
\label{sec:ue_transfer_logic}

In the reference implementation used in this paper, 
once the capacity cell, $c$, decides to shutdown, 
it instructs its \ac{RRC} connected \acp{UE} to perform an A4 inter-frequency handover to the frequency of its paired coverage cell, $b(c)$.
If such \acp{UE} are able to handover within a given predefined time frame (usually of tens of seconds),
the capacity cell, $c$, shuts down.
Otherwise, it abandons its intention to shutdown. 

The A4 inter-frequency handover is triggered when the biased \ac{RSRP}, $M_{n}$, of a neighbouring cell, $n$, of cell, $i$, becomes better than a threshold, $\tau_i^{\rm A4}$,
i.e. $M_{n} + O_{i,n}^{\rm freq} + O_{i,n}^{\rm cell} - H_i^{\rm A4} > \tau_i^{\rm A4}$.
Specifically, according to this expression, it should be noted that, 
in the A4 inter-frequency handover entry condition,
a hysteresis, $H_i^{\rm A4}$, a cell-individual offset, $O_{i,n}^{\rm cell}$, and a cell-specific frequency offset, $O_{i,n}^{\rm freq}$, are used to avoid ping-pongs,
and prioritize a given neighboring cell or a given frequency, respectively.

Accordingly, the optimization of the A4 inter-frequency handover parameters also plays a role in network energy efficiency, 
and are hence part of our framework. 
\section{Network Energy Efficiency Modelling}
\label{sec:modellingChallengesAndSRCON}

As can be inferred from previous sections,
the network energy efficiency modelling problem deals with multiple objectives 
(e.g. {\color {black} network energy consumption and \ac{UE} throughput}) 
and optimization variables 
(e.g. carrier shutdown and A4 handover parameters),
and is particularly challenging due to its large-scale, stochasticity and non-stationarity as well as the complex coupling between cells and the intricate trade-offs between energy consumption and \ac{UE} performance.
For instance, the shutdown of a cell may save energy, 
but impacts, 
not only the performance of the \acp{UE} connected to such cell,
but also the overall network coverage and the performance of those \acp{UE} connected to the neighboring cells as well as the possibility of nearby capacity cells to shutdown.
In practice,
these challenges make the utilization of precise networking models essential to perform a rigorous network energy efficiency optimization.

Fortunately,
recent advancements in big data acquisition and processing have made possible to efficiently store ---and subsequently process--- the large amount of radio measurements to which \acp{BS} have access,
opening the door to new, data-driven modelling and optimization paradigms~\cite{Dai2019,3GPPTR37.816}.

The main idea behind solving the network energy efficiency optimization problem in a data-driven manner is to leverage such easily accessible measurements to generate accurate network-specific models,
without the need of resorting to expensive expert knowledge, \ac{DT}-based data or incomplete \ac{3D} high-definition maps for ray-tracing purposes.

Data-driven network modelling helps addressing most of the challenges of state of the art approaches surveyed in Section \ref{sec:intro},
but it also brings its own. 
Among the challenges introduced by data-driven network modelling,
it is worth highlighting the following two:
\begin{enumerate}
    \item Massive data sets:
    A typical cellular network in a metropolis has around 50~thousands \acp{BS}, each of them generating nearly 3000 \acp{KPI} per hour \cite{KPIs}. When including \ac{UE} measurement reports, this
    results in 1~terabyte of data per hour \cite{SRCON2023Intro}.
    This overwhelming amount of information needs to be stored and processed in time for its productive utilization.
    \item Generalization:
    This is arguably the most relevant issue in data-driven modelling,
    and pertains to the inference of the network performance when applying the model in a scenario --or using a combination of parameters-- never measured before
    and thus not observable in the training data. 
\end{enumerate}

To address these challenges, 
\ac{SRCON} ---our proposed modelling approach---
combines a variety of wireless as well as data-driven and \ac{ML} concepts.
Rather than attempting to replicate network behaviours at a time-scale of milliseconds,
which is fundamentally infeasible, 
considering the aforementioned modelling challenges in practical networks,
\ac{SRCON} emulates network behaviours in a \emph{statistically indistinguishable} manner.
This statistical indistinguishability lays the ground for a practical and accurate overall assessment of \ac{UE} performance,
allowing the use of sufficient statistics ---instead of all available samples--- through an efficient processing of massive data sets.
To enable an accurate generalization, 
and deal with the heterogeneity and complexity of devices,
\ac{SRCON} uses measurement data from live networks to jointly fit \ac{ML}-based, black-box and expert-based, white-box models,
as hinted earlier.
For our network energy efficiency modelling problem,
in more details,
we propose to model \ac{BS} energy consumption and \ac{UE} throughput statistics using customised \ac{ML}-based modeling approaches to the available data to capture the particularities of the specific \ac{BS} products and off-the-shelf \acp{UE} in the area of study 
(e.g. not all \ac{BS} product versions have the same energy consumption characteristics, 
not all \acp{UE} have the same decoding capabilities).
A systematic feature importance analysis was used to identify the most relevant input features of such two \ac{ML}-based models,
e.g. the amount of time for which capacity cells will be in carrier shutdown, the number of \acp{UE} transferred to each neighbouring cell when carrier shutdown takes place, the resulting \ac{PRB} load in those neighbouring cells.
To allow generalisation,
a customised expert-based model is used to derive the inputs of the \ac{ML}-based models for any carrier shutdown and A4 handover parameter configuration.
In other words,
said  white-box model takes as input, among others, the carrier shutdown and A4 handover parameter configuration, 
and provides as output the inputs of our black-box models.
In this manner, we can drive universal ML-based \ac{BS} energy consumption and \ac{UE} throughput predictions,
and further derive network energy efficiency. 

\begin{figure*}[t!]
    \centering
    \includegraphics[trim={3.9cm 6.4cm 3.8cm 6.6cm},clip,scale=0.7]
    {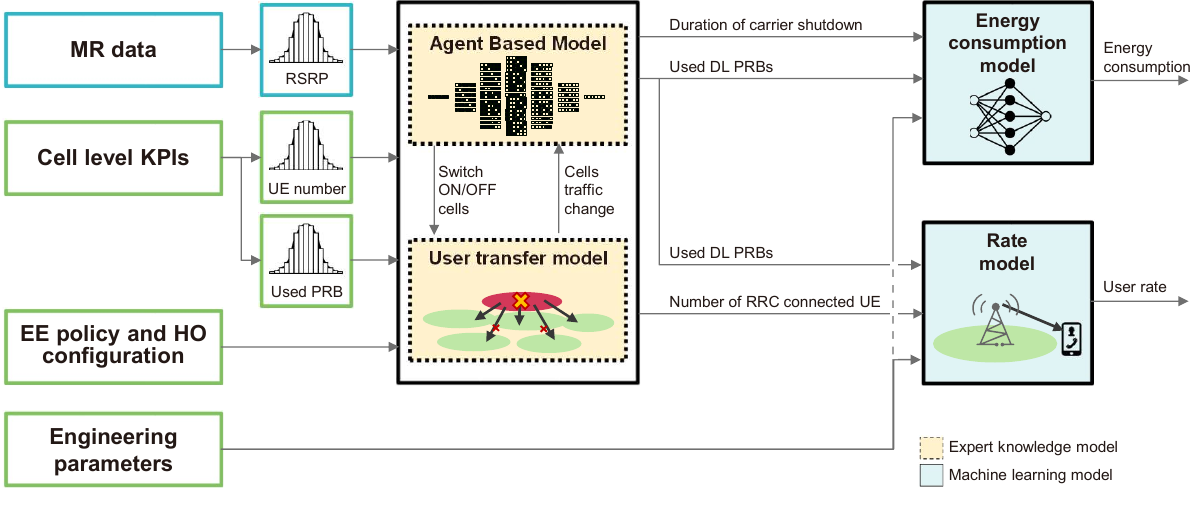}
    \caption{\textcolor{black}{\ac{SRCON} framework for network energy efficiency.}}
    \label{fig:model_architecture}
\end{figure*}

Fig.~\ref{fig:model_architecture} present this framework,
which is further detailed in the following sections. 

For the sake of space, 
note that, in general, we do not mention \ac{UL} related statistics in the rest of the paper, 
but whenever we refer to a \ac{DL} statistic or process, 
the analogous \ac{UL} one is generally implied. 
\section{Modelling objective and data available}
\label{sec:data}



To estimate the goodness of a network optimization campaign, 
the performance of the network is usually measured before and after a change of network parameters. 
Embracing this methodology,
at least two measurement campaigns are usually conducted to collect data from the network when carrying a network optimization exercise. 
During the first measurement campaign, 
before the optimization,
all energy saving solutions, including carrier shutdown, are deactivated to estimate the baseline network energy efficiency and performance. 
The parameters of the network, also refereed to as engineering parameters, are recorded too.  
In the context of this paper, 
we will refer to the data collected in this phase as \emph{unbiased data},
as it is not affected/biased by any energy saving policy. 
During the second measurement campaign, 
after the optimization,
new measurements are collected to estimate the network energy efficiency and performance resulting from the activation of carrier shutdown when using the optimized carrier shutdown and A4 handover parameter configurations.
The new engineering parameters of the network are also recorded.

To develop our new, data-driven modelling framework, 
we take advantage of several such campaigns in different cities and at different times of the year. 
In a nutshell, 
our modelling objective is to predict, 
using the unbiased data of the first measurement campaign, 
the network energy efficiency and performance with the minimum possible error under the carrier shutdown and A4 handover parameter configuration in the second measurement campaign. 
The corresponding metrics to assess the quality of the prediction are generally the \ac{MAE} and \ac{MAPE}.
Importantly, 
it should be noted that the quality of the model is impacted by the amount of data features collected, the amount of samples collected per feature and their time granularity. 

Given the use of \ac{ML} modeling tools in our work,
the data of the second measurement campaign is divided into two sets,
one used for training and validating such \ac{ML} models,
as it will be shown in Sections~\ref{sec:pc_model} and~\ref{sec:rate_model}, and the other for test the overall model, as it will be discussed in Section~\ref{sec:abm_model}.


In the rest of this section, 
we describe the data sets and features available and used in our modelling exercise,
which can be categorised into three types, 
i.e. engineering parameters, cell-level \acp{KPI} and \ac{UE} measurement reports.
We also formally present the notation used in this paper.
Importantly, 
to make sure that this modelling framework can be widely used,
we target at solely using data readily available to any operator.

\subsection{Engineering parameters}
\label{sec:engineering_parameters}

The engineering parameters data sets, $d^{\rm EP}$, describe the network configuration. 
We should distinguish 3 types of engineering parameter data sets: 
\begin{itemize}
    \item \textit{Network parameters data set, $d^{\rm EP}_{\rm N}$}: 
    Information related to the configuration of each \ac{BS}, radio unit and cell
    (e.g. type of radio unit, location, number of \ac{RF} chains, number of supported and configured carriers per radio unit, as well as frequency, bandwidth, bearing, tilt and other information per cell);
    \item \textit{Mobility parameters data set, $d^{\rm EP}_{\rm M}$}: 
    Information related to the configuration of handover procedures of every cell
    (e.g. A4 handover parameters);    
    \item \textit{Energy saving parameters data set, $d^{\rm EP}_{\rm ES}$}: 
    Information related to the configuration of energy saving solutions of every cell
    (e.g. capacity and coverage cell pairing, carrier shutdown thresholds).  
\end{itemize}

Table \ref{tab:eng_param} details the most relevant engineering parameters used in our modelling framework,
and provides a formal definition. 
 
\begin{table}[!t]
\centering
\caption{Engineering parameters}
\label{tab:eng_param}
\begin{tabular}{p{0.1\columnwidth}p{0.8\columnwidth}}
\toprule
\midrule
\multicolumn{2}{c}{\textbf{Network parameters}}   \\
Parameter  & Definition\\
\midrule
${\cal RU}$ &  Set of radio units, ${\cal RU} = \{1, \cdots, r, \cdots, N_{\mathrm{RU}}\}$    \\
$N_{\mathrm{RU}}$   &  Number of radio units    \\
${\cal C}$ &  Set of cells, ${\cal C} = \{1, \cdots, i, \cdots, N\}$    \\
$N$   &  Number of cells    \\
${\cal C}_r$   &  Set of cells operated by radio unit, $r$    \\
$f_i$   &  Carrier frequency of cell, $i$    \\ 
$B_i$   &  Bandwidth of cell, $i$    \\ 
$P_i^{\rm max}$   &  Maximum transmit power of cell, $i$    \\ 
$N_i^{\rm DL}$   &  Number of available \ac{DL} PRBs in cell, $i$    \\  
$N_i^{\rm UL}$   &  Number of available \ac{UL} PRBs in cell, $i$    \\  
\midrule
\multicolumn{2}{c}{\textbf{Mobility parameters}} \\
Parameter & Definition\\
\midrule
$\tau_i^{\rm A4}$       &  A4 handover threshold in cell, $i$ \\ 
$H_i^{\rm A4}$          &  A4 handover hysteresis in cell, $i$  \\
$O_{i,j}^{\rm freq}$    &  A4 handover cell-specific frequency offset in cell, $i$, for the inter-frequency neighboring cell, $j$ \\
$O_{i,j}^{\rm cell}$    &  A4 handover cell-individual offset in cell, $i$, for the inter-frequency neighboring cell, $j$    \\
\midrule
\multicolumn{2}{c}{\textbf{Energy saving parameters}} \\
Parameter & Definition\\
\midrule
${\cal C_{\rm C}}$      &  Set of capacity cells, ${\cal C_{\rm C}} = \{1, \cdots, c,\cdots,  N_{\rm C}\}$    \\
$C$                     &  Number of capacity cells    \\
${\cal C_{\rm B}}$      &  Set of coverage cells, ${\cal C_{\rm B}} = \{1,\cdots,  j,\cdots,  N_{\rm B}\}$    \\
$B$                     &  Number of coverage cells    \\
$b(c)$                  &  Paired coverage cell of capacity cell, $c$ \\
\multicolumn{2}{l}{Carrier shutdown entry conditions} \\
$\chi^{\rm UE}_{c}$     &  RRC connected UE threshold in capacity cell, $c$ \\
$\chi^{\rm DL}_{c,b(c)}$   &  \ac{DL} PRB threshold in capacity cell, $c$, w.r.t. its paired coverage cell, $b(c)$ \\ 
$\chi^{\rm UL}_{c,b(c)}$   &  \ac{UL} PRB threshold in capacity cell, $c$, w.r.t. its paired coverage cell, $b(c)$  \\ 
\multicolumn{2}{l}{Carrier shutdown leaving conditions} \\
$\Psi^{\rm UE}_{c}$     &  RRC connected UE threshold in capacity cell, $c$   \\
$\Psi^{\rm DL}_{c,b(c)}$   &  \ac{DL} PRB threshold in capacity cell, $c$, w.r.t. its paired coverage cell, $b(c)$      \\ 
$\Psi^{\rm UL}_{c,b(c)}$   &  \ac{UL} PRB threshold in capacity cell, $c$, w.r.t. its paired coverage cell, $b(c)$  \\
\bottomrule
\end{tabular}
\end{table}

\subsection{Cell-level KPIs}
\label{sec:cell-level_KPIs}

The cell-level \acp{KPI} data set, $d^{\rm KPI}$, describes the performance of each cell in the network~\cite{3GPPTS32.450}.
Importantly, it should be noted that, to conserve memory at the \ac{BS}, 
this cell-level \acp{KPI} information is typically aggregated over configurable periods of 5, 15, 30, or 60 minutes using sums or averages. 
As a result, 
it does not provide an accurate understanding of the network behavior at the subframe or slot level.
We can distinguish among three types of cell-level \acp{KPI} data sets:
\begin{itemize}
    \item \textit{Traffic statistics data set, $d^{\rm KPI}_{\rm T}$}: 
    Information on the serviced traffic per cell 
    (e.g., average number of active \acp{UE} per \ac{TTI}, average number of used \acp{PRB} per \ac{TTI}, sum traffic volume);
    \item \textit{Energy saving statistics data set, $d^{\rm KPI}_{\rm ES}$}: 
    Information on the activated energy saving modes per cell
    (e.g., duration of the carrier shutdown activation);
    \item \textit{Energy consumption statistics data set, $d^{\rm KPI}_{\rm EC}$}: 
    Information on the energy consumed by each radio unit.
\end{itemize}

Table \ref{tab:Cell-level_KPIs} details the most relevant cell-level \acp{KPI} used in our modelling framework,
and provides a formal definition. 
Note that the level of aggregation was 60 minutes in our cell-level \acp{KPI} data sets, 
and thus we have one entry in each cell-level \acp{KPI} data set per cell, $i$, and hour, $h$. 

\begin{table}[!t]
\centering
\caption{Cell-level KPIs}
\label{tab:Cell-level_KPIs}
\begin{tabular}{p{0.1\columnwidth}p{0.8\columnwidth}}
\toprule
\midrule
\multicolumn{2}{c}{\textbf{Traffic statistics}}   \\
Parameter  & Definition\\
\midrule
$U^{\rm UE}_{i,h}$  & average number of RRC connected \acp{UE} in the cell, $i$, at hour, $h$ \\
$\Delta^{\rm DL}_{i,h}$  & average number of \ac{DL} PRBs used per \ac{TTI} in cell, $i$, at hour, $h$    \\
$\Delta^{\rm UL}_{i,h}$  & average number of \ac{UL} PRBs used per \ac{TTI} in cell, $i$, at hour, $h$    \\  
$V^{\rm DL}_{i,h}$  & number of \ac{DL} bits successfully transmitted at the \ac{RLC} layer in cell, $i$, at hour, $h$    \\  
$V^{\rm DL-}_{i,h}$ & number of \ac{DL} bits successfully transmitted in the last time slots during which the buffer becomes empty at the \ac{RLC} layer in cell, $i$, at hour, $h$    \\  
$V^{\rm UL}_{i,h}$  & number of \ac{UL} bits successfully received at the \ac{RLC} layer in cell, $i$, at hour, $h$    \\ 
$T^{\rm DL}_{i,h}$  & amount of time in which cell, $i$, was transmitting \ac{DL} bits at hour, $h$ \\    
$T^{\rm DL-}_{i,h}$ & amount of time in which cell, $i$, was transmitting \ac{DL} bits at hour, $h$, excluding the last time slots during which the \ac{DL} buffer becomes empty \\  
$T^{\rm UL}_{i,h}$  & amount of time in which cell, $i$, was receiving \ac{UL} bits at hour, $h$     \\
$RN^{\rm DL}_{g,i,h}$ & number of samples with the \ac{UE} rate falling into the $g$-th predefined \ac{UE} rate range in cell, $i$, at hour, $h$,
where $g \in \{1, \cdots, 15\}$\\

\midrule
\multicolumn{2}{c}{\textbf{Energy consumption statistics}} \\
Parameter & Definition\\
\midrule
$t^{\rm CS}_{i,h}$    & duration of carrier shutdown in capacity cell, $c$, at hour $h$ \\
$E^{\rm RU}_{r,h}$    & energy consumption of radio unit, $r$, at hour $h$    \\ 
\bottomrule
\end{tabular}
\end{table}

\subsection{UE measurement reports}
\label{sec:measurement_reports}

The \ac{UE} can be directed to perform a variety of measurements, 
including intra-frequency, inter-frequency, and inter-\ac{RAT} measurements, 
in accordance with the measurement configuration provided by the network~\cite{3GPPTS36.331}.
This valuable information gathered at the \ac{UE} is transmitted to the serving \ac{BS} through \ac{UE} measurement reports.
Each \ac{UE} measurement report is uniquely identified by the \ac{ID} of the \ac{UE} that conducted the measurement and the time at which the measurement was taken.
Among other statistics, 
these reports include the \ac{RSRP} measured by the \ac{UE} from its serving cell, as well as those from a limited number of neighboring cells.

Our data set of measurement reports, denoted as $d^{\rm MR}$, comprises the compilation of all measurement reports received by all cells in the network from their respective \acp{UE} during specific hours of the day.
Table \ref{tab:UE_measurement_reports} provides a comprehensive breakdown of the most pertinent features present in a \ac{UE} measurement report as used in our modeling framework, 
accompanied by a formal definition.
This data set contains one entry per measurement report.

Considering that measurement reports can be transmitted by \acp{UE} in connected mode as frequently as every 5 milliseconds, 
the measurement report data set, $d^{\rm MR}$, occupies significantly more memory than the engineering parameters, $d^{\rm EP}$, and the cell-level \acp{KPI}, $d^{\rm KPI}$, data sets.
As a result, measurement reports are usually collected and stored for shorter durations, 
often just a few hours during specific time periods.

\begin{table}[!t]
\centering
\caption{UE measurement reports}
\label{tab:UE_measurement_reports}
\begin{tabular}{p{0.1\columnwidth}p{0.8\columnwidth}}
\toprule
\midrule
\multicolumn{2}{c}{\textbf{UE measurement report statistics}}   \\
Parameter  & Definition\\
\midrule
$T$                & Timestamp of measurement \\
$ID_{u}$           & ID of \ac{UE}, $u$ \\
$S^{\rm ID}_{u}$   & ID of the serving cell of \ac{UE}, $u$ \\
${\cal N}^{\rm ID}_{u}$      & Set of IDs of the neighbouring cells of \ac{UE}, $u$ \\
$M_{u,i}$          & \ac{RSRP} of neighbouring cell, $i \in  S^{\rm ID}_{u} \cup {\cal N}^{\rm ID}_{u}$, measured by of \ac{UE}, $u$ \\
\bottomrule
\end{tabular}
\end{table}
\section{ML-based radio unit energy consumption model}
\label{sec:pc_model}

Understanding and optimizing energy-saving aspects necessitates a thorough grasp of \ac{RAN} energy consumption modeling. Accurately approximating \ac{RAN} energy consumption involves summing the energy expended by its radio and base band units. 
In this section, we outline the framework underpinning our radio unit energy consumption model, 
centered around \ac{ML}. 
Notably, an \ac{ANN} architecture emerged as our choice to model the energy consumption, $E^{\rm RU}_{r,h}$, per radio unit, $r$, at hour, $h$, after rigorous exploration of various \ac{ML} models due to its robust performance and broad applicability. 
While the selection process is not presented here for brevity, 
a comprehensive account can be found in~\cite{piovesan2022machine} for further insights into this radio unit energy consumption model and our choices.

\subsection{Feature importance analysis}
\label{sec:PC-fia}

To discern the most impactful factors in estimating radio unit energy consumption, 
a feature importance analysis was conducted using data obtained over a span of 12 days from an extensive deployment featuring 7500 \ac{4G}/\ac{5G} radio units. 
This encompassed 24 distinct types of radio units (commercial products). 
The data sets utilized in this analysis included engineering parameters, $d^{\rm EP}$, and cell-level \acp{KPI}, $d^{\rm KPI}$, from the measurement campaign. 
These data sets comprised 150 features per cell, 
with the most pertinent ones highlighted in Table~\ref{tab:eng_param} and~\ref{tab:Cell-level_KPIs}. 
It is worth noting that during the data collection, 
energy-saving measures such as symbol, channel, and carrier shutdown were active. 
Additionally, as indicated in Section~\ref{sec:cell-level_KPIs}, 
the cell-level \acp{KPI} statistics were recorded on an hourly basis.

The feature importance analysis was conducted in two phases:
\emph{i)} 
Gradient boosting models were trained with varying sets of input features;
\emph{ii)} 
Then,
an examination of the \ac{SHAP} values for each feature was undertaken using these models.
To elaborate, 
the \ac{SHAP} value assigned to each feature signifies the alteration in the model's anticipated prediction when that specific feature is taken into account~\cite{NIPS2017_7062}.

\begin{figure}[!t]
    \centering
    \includegraphics[scale=0.35]{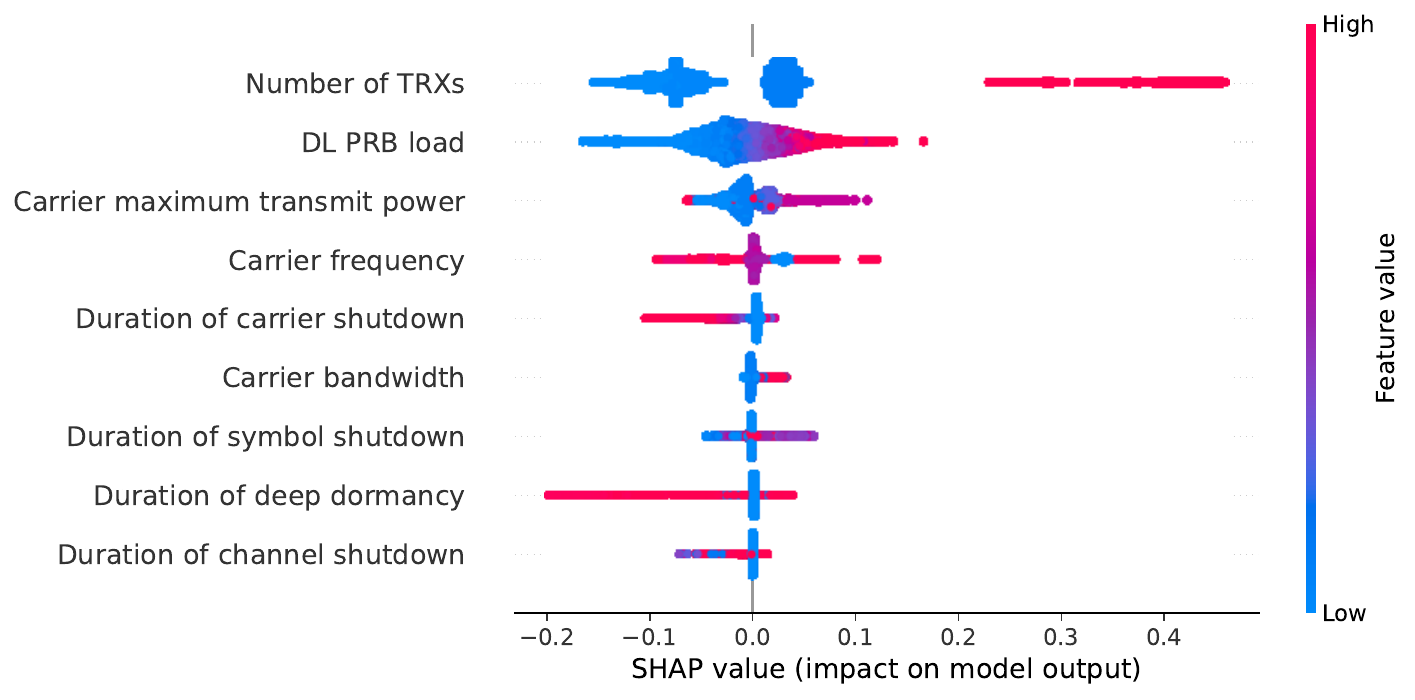}
    \caption{Energy consumption SHAP analysis performed on the most important numerical features in the collected measurements data.}
    \label{fig:shap}
\end{figure}

Fig.~\ref{fig:shap} illustrates the \ac{SHAP} values associated with the key numerical features in the data set.
Specifically, 
the diagram illustrates both the magnitude and direction of each feature's impact on the model output in relation to the average model prediction.
The right-side y-axis denotes the corresponding feature value, displayed in a color gradient from blue (low values) to red (high values).
Each individual scatter dot corresponds to a data instance.

Our feature importance analysis revealed that the \ac{DL} PRB load, 
represented as $\frac{\Delta^{\rm DL}_{i,h}}{N_i^{\rm DL}}$, 
holds the highest significance in modeling radio unit energy consumption. 
Additionally, the second most crucial feature is the maximum transmit power denoted by $P_i^{\rm max}$. 
These two features enable the model to effectively capture the power transmitted, $P^{\rm TX}_{i}$, by cell, $i$, across various \ac{DL} \ac{PRB} load levels. 
The duration of carrier shutdown activation, $t^{\rm CS}_{i,h}$, also emerges as significant, 
offering insights into the sleep behavior of the cells.
The estimation of radio unit energy consumption is also influenced by factors such as the radio unit type, the number of \acp{TRX}, carrier transmission mode, frequency, and bandwidth. 
These parameters define the radio unit's hardware and capabilities, 
making them crucial for accurate modeling.
A notable finding from our analysis is the strong correlation between the \ac{MCS} and the number of \acs{MIMO} layers used per \ac{DL} \ac{PRB} and the \ac{DL} \ac{PRB} load. 
This implies that modeling the \ac{DL} \ac{PRB} load across all cells operated by a radio unit may be sufficient to capture their average energy consumption behavior.

Table~\ref{tab:ML_inputs} provides a comprehensive list of the 10 selected features following the feature importance analysis. 
Features that have minimal impact on energy consumption or are highly correlated with the chosen ones, 
thus offering limited additional information, 
have been excluded.

\begin{table}[!t]
\caption{Energy consumption model input parameters.}
\centering
\begin{tabular}{@{}lll@{}}
\toprule
Class                    & Parameter                        & Type               \\ \midrule
Engineering parameter    & Radio type                       & Categorical        \\
Engineering parameter    & Number of  TRXs                  & Numerical          \\
Engineering parameter    & Carrier transmission mode        & Categorical \\
Engineering parameter    & Carrier frequency                & Numerical          \\
Engineering parameter    & Carrier bandwidth                & Numerical          \\
Engineering parameter    & Carrier maximum transmit power   & Numerical \\
Traffic statistics       & Carrier DL PRB load              & Numerical          \\
Energy saving statistics & Duration of carrier shutdown     & Numerical          \\ \bottomrule
\end{tabular}
\label{tab:ML_inputs}
\end{table}

\begin{rmk}
While the maximum transmit power, $P_i^{\rm max}$, of cell, $i$, and other selected inputs in Table~\ref{tab:ML_inputs} are static variables provided by the engineering parameters, 
the \ac{DL} \ac{PRB} load, $\frac{\Delta^{\rm DL}_{i,h}}{N_i^{\rm DL}}$, and the duration, $t^{\rm CS}_{i,h}$, of carrier shutdown depend on the carrier shutdown and A4 handover parameter settings. 
Therefore, they need to be modeled as functions of these parameters to accurately predict network energy consumption for different configurations, 
enabling effective optimization.
\end{rmk}

Section~\ref{sec:abm_model} presents the white-box model that allows such generalization.

\subsection{Inputs of the model}
\label{sec:PC-inputlayer}

Each of the input features listed in Table~\ref{tab:ML_inputs} underwent pre-processing based on its type to eliminate outliers, 
and was then fed into the \ac{ANN}. 
The numerical features were normalized prior to entering the model, 
while the categorical features were encoded using one-hot encoding.

To ensure maximum generality and flexibility, 
our \ac{ANN} model takes input data from $C^{\mathrm{MAX}}$ carriers, 
where $C^{\mathrm{MAX}}$ represents the highest number of carriers that the most capable radio unit, 
can manage. 
In our data set, $C^{\mathrm{MAX}}$ is six. 
When a radio unit handles fewer carriers, $C<C^{\mathrm{MAX}}$,
the input neurons corresponding to the remaining $C^{\mathrm{MAX}}-C$ carriers are assigned zero values. 
This universal model approach enables the implementation of a unique \ac{ANN} model with a fixed number of input neurons. 
It can be trained using data from all radio units, $\forall r \in {\cal RU}$, in the data set, 
regardless of their number of configured carriers, $|{\cal C}_r|$, 
while resulting in minimal loss in accuracy, 
as it will be discussed in Section~\ref{sec:PC-results}.

\subsection{Outputs of the model}
\label{sec:PC-outputlayer}

The analysis of the collected data revealed instances where different energy consumption values, $E^{\rm RU}_{r,h}$, are reported for the same input feature values. 
This variability can be attributed to several factors, 
including:
\begin{enumerate}
\item Features that have a slight impact on energy consumption but are not included in our data sets.
\item Potential errors in measurements or data collection.
\item Tolerances of hardware components, which influence their energy consumption behavior.
\end{enumerate}
For the sake of presentation,
we use the following change of notation, $\bar{y} = E^{\rm RU}_{r,h}$, in this section.

To characterize this noise, 
we define the measured energy consumption, $\bar{y}$, as $\bar{y} = y + n$, 
where $y$ represents the energy consumption for a given input configuration, 
and $n$ accounts for the noise arising from the aforementioned factors. 
Based on our data analysis, 
we deduced that the noise term $n$ can be approximated as a normally distributed variable with a mean of zero and a standard deviation of $\sigma$. 
Consequently, the measured energy consumption $\bar{y}$ follows a normal distribution with a mean of $\mu=\mathbb{E}[\bar{y}]$ and a standard deviation of $\sigma$.

To account for this noise, 
our \ac{ANN} model was designed to estimate and output both of these parameters, $\mu$ and $\sigma$, for a given input, $x$. Importantly, 
the outputs of these two parameters enable the computation of a confidence interval for each energy consumption estimate,
enhancing the overall trustworthiness of the estimation process.

\subsection{Architecture of the model}
\label{sec:PC-arch}

\begin{figure} [!t]
    \centering
    \includegraphics[scale=0.6]{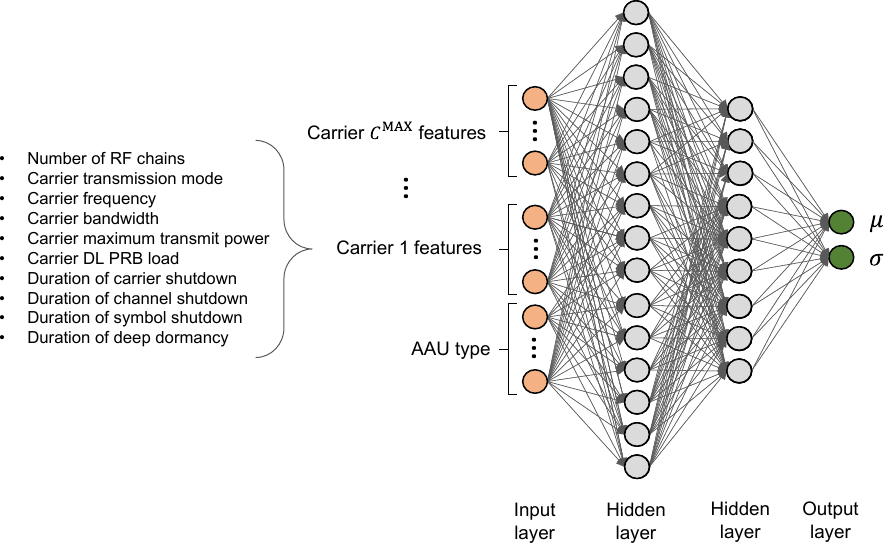}
    \caption{{\color {black}Structure of our proposed ANN, highlighting the selected input variables, the organization and interconnection of the two hidden layers, and the configuration of the two output neurons which represent the mean and standard deviation of the power consumption estimates.}}
    \label{fig:MLmodel}
\end{figure}

The fundamental architecture chosen for our proposed \ac{ANN} model is the multilayer perceptron, 
which comprises several fully connected layers of neurons~\cite{Goodfellow-et-al-2016}.
{\color {black} Fig.~\ref{fig:MLmodel} offers a comprehensive visualization of the overall \ac{ANN} employed for estimating power consumption in this paper.}  

{\color {black} Following from the figure}, 
the input layer comprised $n_{\rm I}=N_{\mathrm{RU}}+10*C^{\mathrm{MAX}}$ neurons, 
where $N_{\mathrm{RU}}$ represents the number of distinct radio unit types modeled by the \ac{ANN}. 
In our specific data set, 
$N_{\mathrm{RU}}$ equated to 24, 
and the number of features considered per radio type was 10,
resulting in an input layer of $n_{\mathrm{I}}=84$ neurons. 
This architecture configuration allowed for effective representation of input data.

Two hidden layers followed the input layer, 
with $n_{\rm H,1}=40$ and $n_{\rm H,2}=15$ neurons, respectively. 
These dimensions were determined through an optimization process aimed at maximizing model accuracy.

The output layer consisted of $n_{\rm O}=2$ neurons, 
capturing the mean $\mu$ and standard deviation $\sigma$ of the radio unit energy consumption $\bar{y} = E^{\rm RU}_{r,h}$ for radio unit, $r$, at hour, $h$, 
as explained earlier. 
Given that both metrics are inherently positive, 
a sigmoid activation function was employed at the output layer.

\subsection{Training and testing of the model}
\label{sec:PC-trainningtest}

The objective of the model optimization process was to simultaneously minimize prediction error and uncertainty. Specifically, 
during the training phase, 
the aim was to enhance the probability that energy consumption samples, $\bar{y}$, for a given input, $x$, fall within the estimated distribution, $\mathcal{N}(\mu,\sigma)$. 
This approach ensures that the statistical distribution of energy measurements output by the model aligns with the distribution of energy measurements in the data.

Given that energy consumption, $\bar{y}$, follows a normal distribution, 
this probability was computed as
\begin{equation}
    P\left(\bar{y}|\mu,\sigma\right) = \frac{1}{\sigma\sqrt{2\pi}} e^{-\frac{(\bar{y}-\mu)^2}{2\sigma^2}}.
\end{equation}
To align with optimization methods that aim to minimize,
we adopted the following loss function for training the \ac{ANN} model:
\begin{equation}
    l(\bar{y},\mu,\sigma) = - \log \left( P(\bar{y}|\mu,\sigma)\right) = \log(\sigma) + \frac{(\bar{y}-\mu)^2}{2\sigma^2}.
    \label{eq:loss}
\end{equation}
This loss function serves the dual purpose of minimizing both prediction error and associated uncertainty. 
The first term is minimized when the standard deviation, $\sigma$, is low, 
indicating high confidence in the estimation. 
The second term is minimized when the prediction error, $\bar{y}-\mu$, is reduced.

Regarding the training and testing data sets, 
we utilized the same data set as employed in the feature importance analysis (Section \ref{sec:PC-fia}).  
The data samples were chronologically sorted and divided into a training set (80\,\% of samples, the first 10 days) and a testing set (the remaining 20\,\%, the last 2 days).
For training the ANN model, 
80\,\% of the training set samples were randomly chosen, 
with the remaining 20\,\% serving for model validation during training. 
The Adam version of the gradient descent algorithm was employed for model training~\cite{Goodfellow-et-al-2016}, 
incorporating an early stopping mechanism to halt training after 200 epochs without validation loss improvement.

\subsection{Performance of the model}
\label{sec:PC-results}

To evaluate performance, 
we conducted a comparison between the estimated energy consumption and the actual measurements from the test set, utilizing the \ac{MAE} and \ac{MAPE} as evaluation metrics.

It is important to mention that the training of the \ac{ANN} model took approximately 75 minutes \textcolor{black}{in a machine powered by a Intel\textregistered $\:$ Core\textsuperscript{TM} i7-9700 CPU @ 3.00 GHz with 32~GB of \ac{RAM}}, 
encompassing 1086 iterations with a learning rate of 0.001.

\begin{figure}[!t]
    \centering
\includegraphics[width=8.7cm]{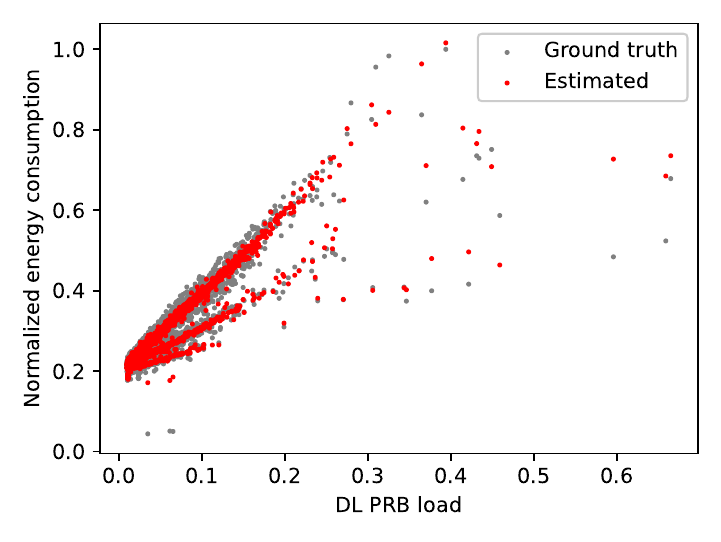}
    \caption{True and estimated normalized energy consumption vs DL PRB load for multiple BSs of a given type.}
    \label{fig:truevsest}
\end{figure}

From the results, 
we can observe that our energy consumption model achieved a \ac{MAE} of 10.94\,W, 
and a remarkably low \ac{MAPE} of 5.87\,\%,
when estimating the energy consumed by each radio unit across all hours of the test period.
As an example,  
Fig.~\ref{fig:truevsest} illustrates the comparison between actual and estimated normalized energy consumption for various radio units of the same type. 
It is worth noting that energy consumption follows a linear relationship with the \ac{DL} \ac{PRB} load, $\frac{\Delta^{\rm DL}_{i,h}}{N_i^{\rm DL}}$, 
and the presence of three distinct slopes is attributed to different configurations of the maximum transmit power, $P_i^{\rm max}$, within the data set. 
The proposed \ac{ANN} model adeptly captures the energy consumption characteristics for each of these configurations.

To complement these results,
it should also be noted that the accuracy loss incurred by our universal \ac{ANN} modelling approach (i.e. one model for all radio units) 
with respect to one in which an \ac{ANN} model is trained per group of radio units supporting the same number of carriers is equal to 1.86\,\%.
This shows that the flexibility of the universal model comes at the expense of a reduced accuracy loss.
\section{ML-based \ac{DL} UE rate model}
\label{sec:rate_model}

In the context of optimizing a large-scale network for minimal energy consumption, 
maintaining an acceptable \ac{UE} rate is crucial to avoid compromising network performance. 
In this section, 
we present a summary of the proposed \ac{ML} model for estimating the \ac{UE} rate, 
while considering the diversity of end-user devices in the network.
It's important to emphasize that while we selected an \ac{ANN} architecture to model the energy consumption, $E^{\rm RU}_{r,h}$, per radio unit, $r$, at hour, $h$, 
we opted for a gradient boosting architecture to model the average, $R^{\rm avg}_{i,h}$, and the 5\%-tile, $R^{\rm ce}_{i,h}$, \ac{UE} rate per cell, $i$, at hour, $h$. 
This choice was based on considerations of both accuracy and complexity.

Unlike energy consumption, 
which remains relatively consistent for the same radio unit operating under similar conditions, 
the \ac{UE} rate is sensitive to the geographical characteristics and channel conditions of a cell's deployment area. Therefore, two cells with identical configurations and traffic loads but deployed in different locations could exhibit vastly different average and 5\%-tile \ac{UE} throughput. 
To capture these nuances effectively, 
our experiments showed that using separate models for individual cells is more appropriate than using a universal model for all cells. 
Furthermore, when pursuing this approach, 
gradient boosting not only yield improved accuracy, 
but also reduced complexity in terms of training time compared to an \ac{ANN} architecture or other alternatives we tested.


\subsection{Feature importance analysis}
\label{sec:rate-fia}

Similar to the approach described in Section~\ref{sec:PC-fia}, 
we conducted a feature importance analysis on the collected data sets to determine the key features for estimating average and 5\%-tile \ac{UE} throughput. 
For consistency,
we also utilized the same data sets as in Section \ref{sec:PC-fia}. 
However, it is important to note that our cell-level \acp{KPI} data sets do not directly provide features corresponding to the average, $R^{\rm avg}_{i,h}$, or 5\%-tile, $R^{\rm ce}_{i,h}$, \ac{UE} throughput for each cell, $i$, at hour, $h$: 
\begin{itemize}
\item
As per~\cite{3GPPTS32.450}, 
the average \ac{UE} rate, $R^{\rm avg}_{i,h}$, for each cell, $i$, at hour, $h$, is calculated by subtracting the number of \ac{DL} bits, $V^{\rm DL-}_{i,h}$, successfully transmitted during the last time slots when the \ac{DL} buffer becomes empty in cell, $i$, at hour, $h$, from the total number of \ac{DL} bits, $V^{\rm DL}_{i,h}$, transmitted at the \ac{RLC} layer at the same cell and hour. 
This value is then divided by the duration, $T^{\rm DL-}_{i,h}$, during which cell, $i$, was transmitting \ac{DL} bits at hour, $h$, 
excluding the last time slots when the \ac{DL} buffer became empty. 
Mathematically, $R^{\rm avg}_{i,h} = \frac{V^{\rm DL}_{i,h} - V^{\rm DL-}_{i,h}}{T^{\rm DL-}_{i,h}}$.
\item
On the other hand, 
the 5\,\%-tile \ac{UE} rate, $R^{\rm ce}_{i,h}$, for each cell, $i$, at hour, $h$, can be estimated from the counters $RN^{\rm DL}_{g,i,h}$ in cell, $i$, at hour, $h$. 
These counters indicate the number of samples falling within a predefined \ac{UE} rate range, $g$.
\end{itemize}

\begin{figure}[t]
    \centering
    \includegraphics[scale=0.4]{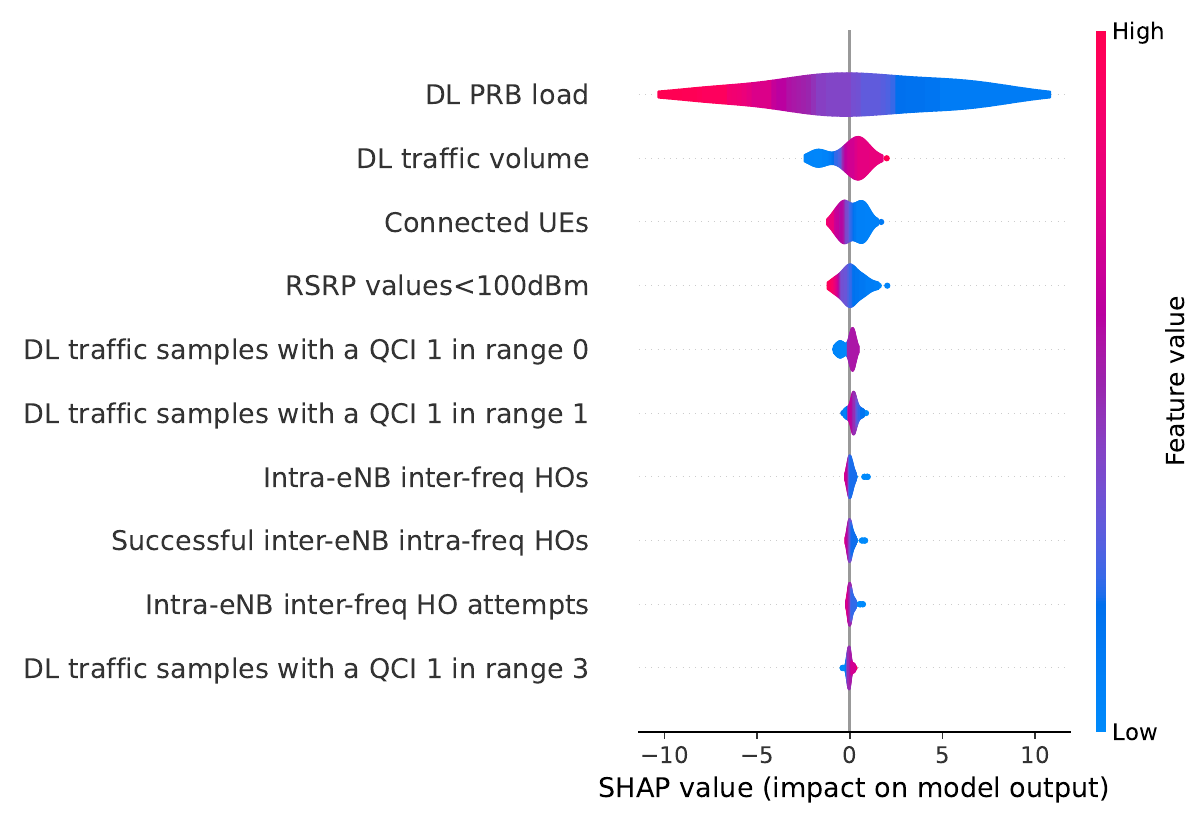}
    \caption{Example of \ac{UE} rate SHAP analysis performed on the most important numerical features in the collected measurements data in a given cell.}
    \label{fig:SHAP_analysis}
\end{figure}

First, we conducted a \ac{SHAP} analysis for each cell in our data set to determine the most relevant features for modeling the \ac{UE} rate per cell and hour~\cite{NIPS2017_7062}. 
An example of the \ac{SHAP} values of the ten most significant features for a specific cell is depicted in Fig. \ref{fig:SHAP_analysis}. 
In this instance, 
the \ac{DL} \ac{PRB} load emerges as the most influential feature on average. 
As expected, 
the \ac{SHAP} analysis indicates that a higher \ac{DL} \ac{PRB} load corresponds to a lower estimated \ac{UE} rate. 
This connection is logical, 
given that a higher \ac{DL} \ac{PRB} load generally implies a greater number of connected \acp{UE}, 
resulting in less bandwidth per \ac{UE}.

\begin{figure*}[ht]
    \centering
    \includegraphics[scale=0.35]{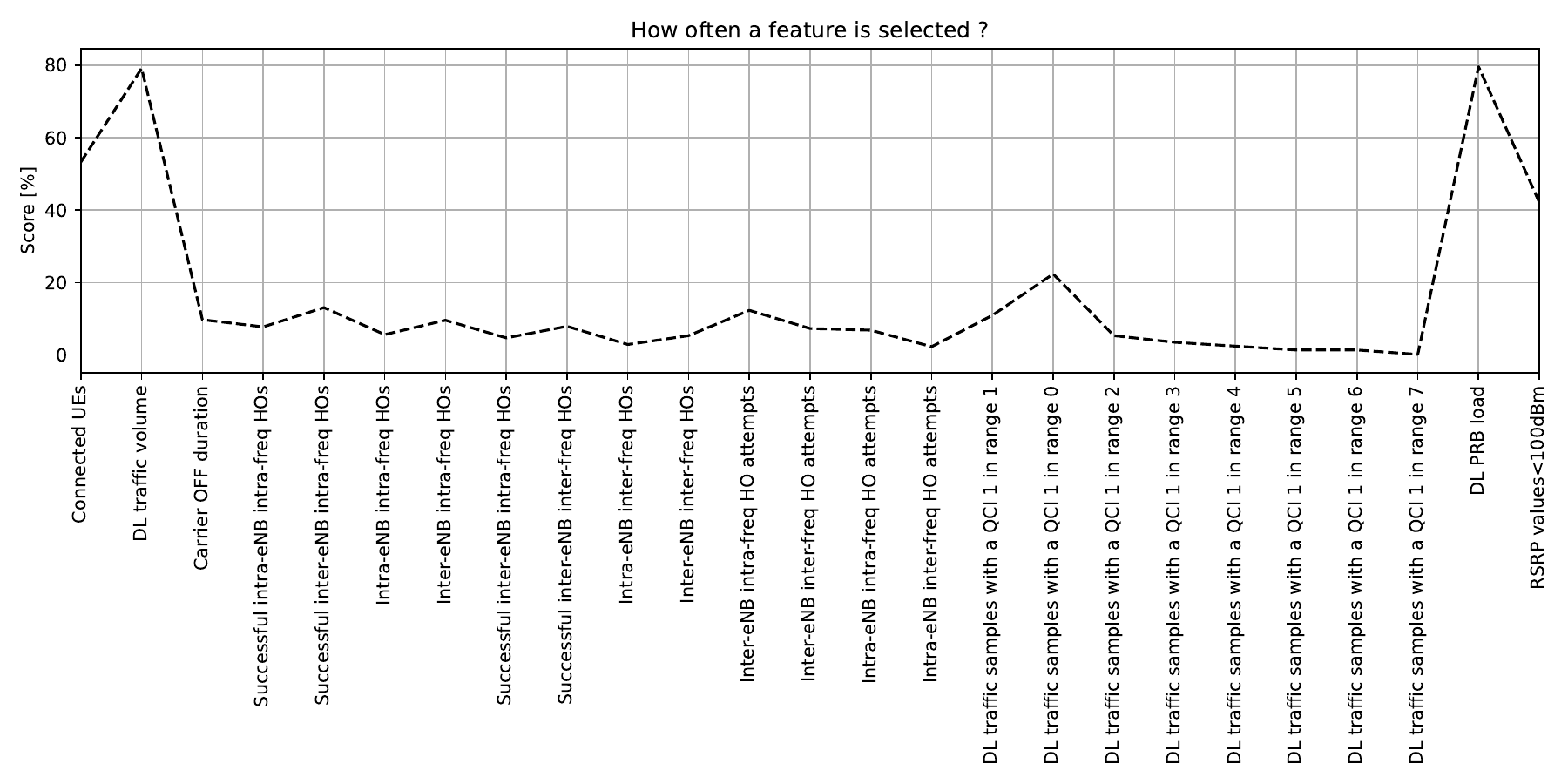}
    \caption{Feature importance analysis for the UE rate model.}
    \label{fig:shap_importance}
\end{figure*}

To provide a comprehensive analysis across all cells, 
we identified the five most contributing features in each cell based on their \ac{SHAP} values and ranked them according to their frequency of occurrence. 
Fig. \ref{fig:shap_importance} illustrates how frequently a feature is included in the five most influential features across all cells and \ac{UE} rate models. 
This graph underscores the significance of the \ac{DL} \ac{PRB} load, \ac{DL} traffic volume, number of \ac{RRC} connected \acp{UE}, and the fraction of \ac{RSRP} samples below $-100$\,dBm in predicting \ac{UE} throughput.

Consequently, 
given the enhanced importance of these features,
as an \textit{ideal baseline estimator}, 
we consider the scenario where all four features are available and utilized to formulate the \ac{UE} rate model for each cell. 
However, in practical implementation, 
due to modelling complexity issue and the correlations among some of these features, 
we construct a model using only two out of the four: 
the \ac{DL} \ac{PRB} load and the number of \ac{RRC} connected \acp{UE}. 
This decision aims to simplify the complexity of our expert-based, white-box \ac{ABM},
which will be elaborated on in Section \ref{sec:abm_model}. 
We will refer to the model that employs only these two aforementioned features as the ``ABM-friendly estimator". 
This section will include a comparison between the ideal baseline estimator and the ABM-friendly estimator.

\begin{rmk} 
The \ac{DL} \ac{PRB} load, $\frac{\Delta^{\rm DL}_{i,h}}{N_i^{\rm DL}}$, and the number, $U^{\rm UE}_{i,h}$, of \ac{RRC} connected \acp{UE} depend on the carrier shutdown and A4 handover parameter settings. 
Therefore, they need to be modeled as functions of these parameters to accurately predict network energy consumption for different configurations, 
enabling effective optimization.
\end{rmk}

\subsection{Inputs and output of the model}

Before constructing our gradient boosting models, 
we conducted pre-processing on the selected input features as discussed in Section \ref{sec:rate-fia}. 
An important insight from our data analysis was that cells with high \ac{DL} PRB load levels tend to have more accurately predicted average and 5\%-tile \ac{UE} throughput.

\begin{figure}[ht]
    \centering
    \adjincludegraphics[height=6cm,trim={{.15\width} {.15\width} {.15\width} {.15\width}},clip]{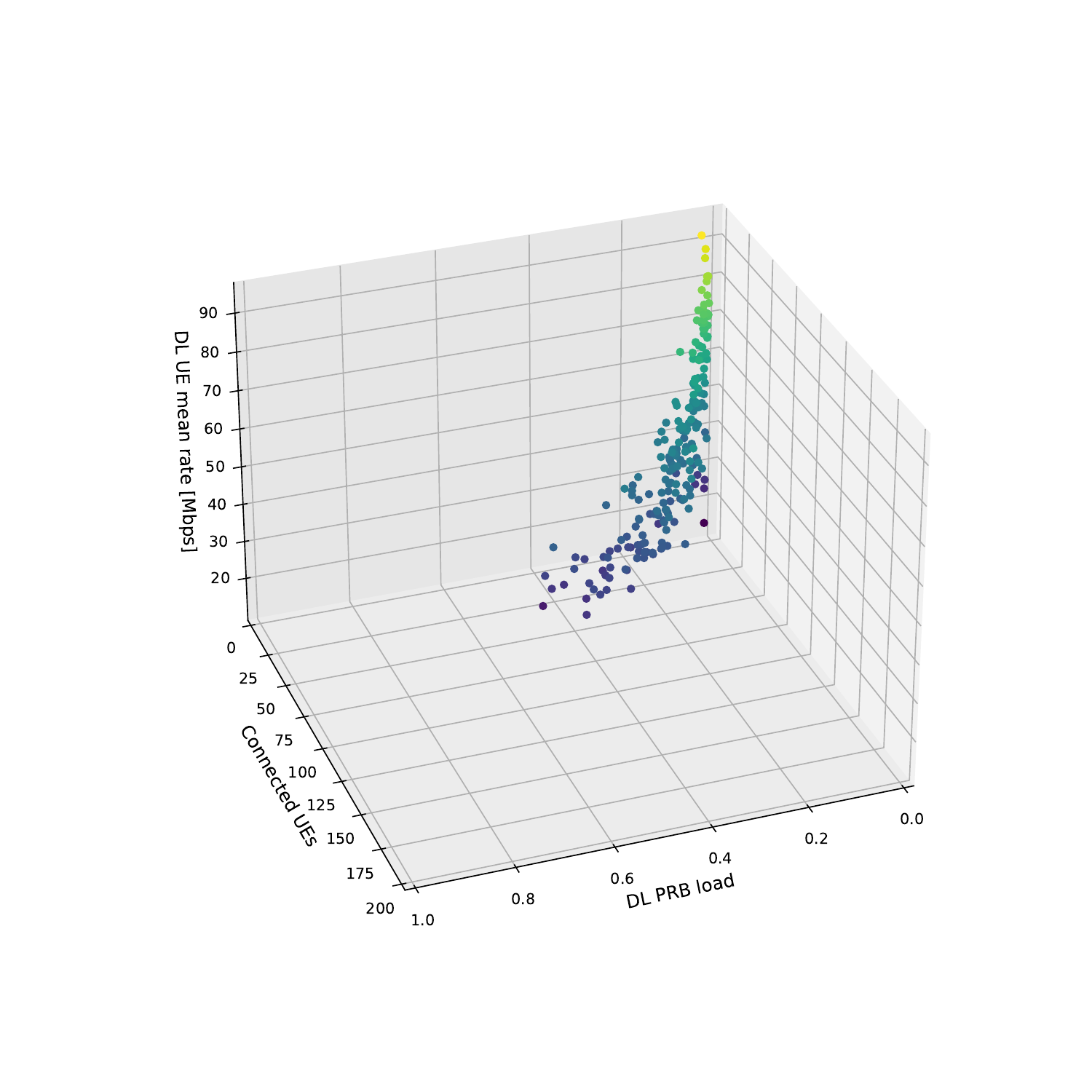}
    \caption{Average \ac{UE} rate for a cell mostly experiencing low loads.}
    \label{fig:lowload}
\end{figure}

\begin{figure}[ht]
    \centering
    \adjincludegraphics[height=6cm,trim={{.15\width} {.15\width} {.15\width} {.15\width}},clip]{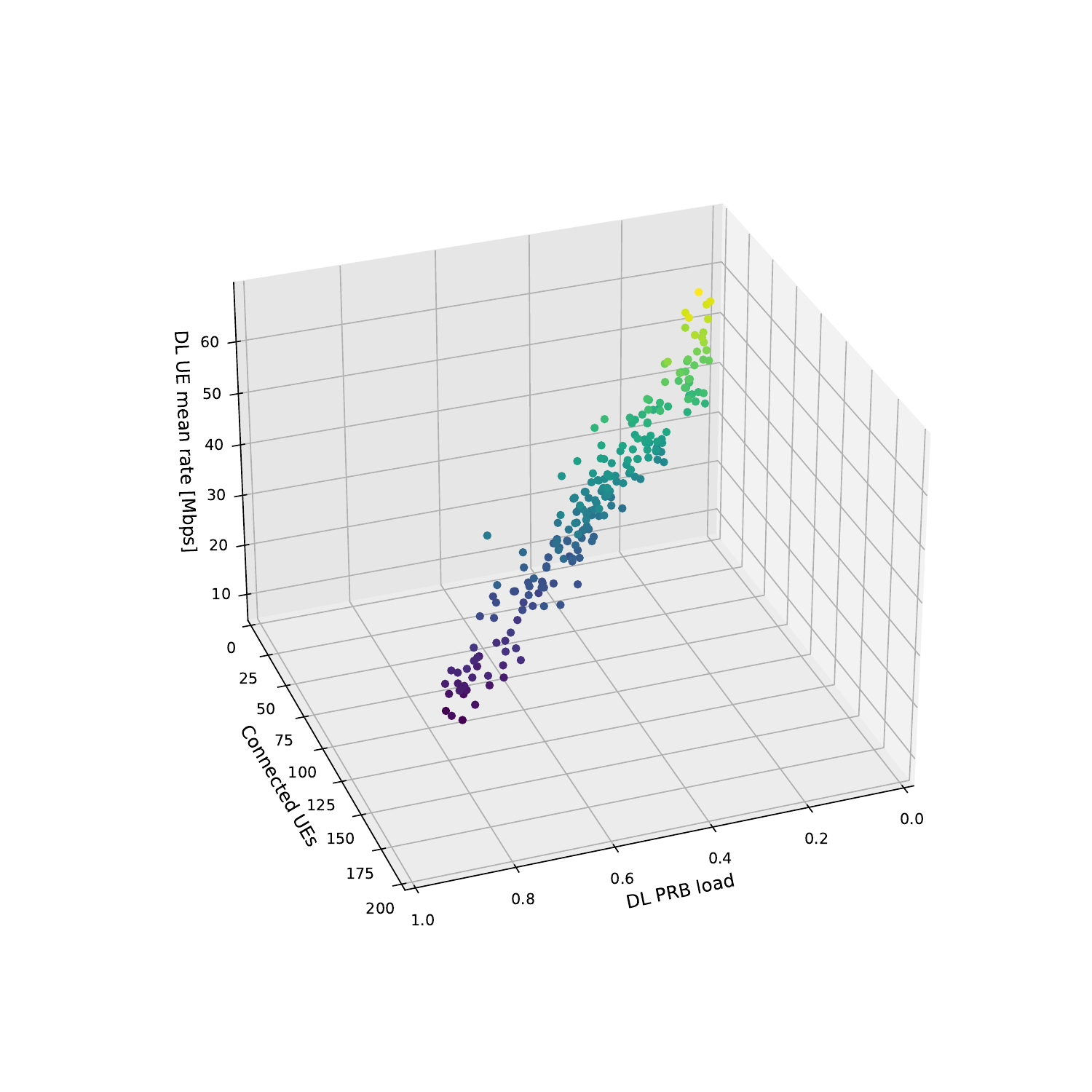}
    \caption{Average \ac{UE} rate for a cell experiencing high and low loads.}
    \label{fig:highload}
\end{figure}

In Fig. \ref{fig:lowload}, 
we observe the average \ac{UE} rate of a cell with a low number of \acp{UE} and consistently low loads, 
while Fig. \ref{fig:highload} depicts a cell with higher \ac{UE} count and larger average loads. 
The former graph highlights the considerable variance in average \ac{UE} throughput when $\frac{\Delta^{\rm DL}_{i,h}}{N_i^{\rm DL}} < 0.1$, 
which impacts the accuracy of \ac{UE} rate estimation using available data.

In intuitive terms, 
when the \ac{DL} PRB load, $\frac{\Delta^{\rm DL}_{i,h}}{N_i^{\rm DL}}$, is low, 
the \acp{UE} sharing the transmission slot might have limited payload, 
potentially leading to underutilized bandwidth. 
Consequently, the average \ac{UE} rate, $R^{\rm avg}_{i,h}$, becomes highly dependent on the specific traffic characteristics, 
such as file sizes. 
For instance, if there are only enough bits to occupy 25\,\% or 75\,\% of the transmission slot's bandwidth, 
the resulting \ac{UE} rate would be 25\,\% or 75\,\%  of the cell's maximum capacity. 
As accurately predicting traffic nature and file sizes based on available data is challenging, 
we decided to model average and 5\%-tile \ac{UE} throughput using our gradient boosting approach, focusing solely on input data corresponding to \ac{DL} PRB load smaller or equal than 0.1, i.e. $\frac{\Delta^{\rm DL}_{i,h}}{N_i^{\rm DL}} \geq 0.1$.
For lower loads, 
they were modeled using a random variable with its \ac{PMF} learned from the data.

\subsection{Architecture and training of the model}

The gradient boosting for regression, 
as developed by \cite{scikit-learn}, 
was employed within this framework. 
This method constructs an additive model using a forward stage-wise approach. 
At each stage, a regression tree is fitted to the negative gradient of the specified loss function.

Given our objective of minimizing prediction errors for a single output, 
we opted for the mean squared error as the chosen loss function.

Regarding the training and testing phases, 
we utilized the same data set as employed in the feature importance analysis (refer to Section \ref{sec:rate-fia}). 
Notably, this data set also aligns with the one used throughout Section \ref{sec:pc_model}. 
We applied the same training and testing methodology as described in Section \ref{sec:PC-trainningtest}, 
thereby yielding consistent training and testing sets. 

The process of model training was executed through the adoption of the Friedman version of the gradient descent algorithm, 
as documented in \cite{hastie_09_elements-of.statistical-learning}. 
Given the manageable time frame allocated for training, 
the technique of early stopping was intentionally omitted.

\subsection{Performance of the model}
\label{res:rate_model}

To evaluate performance, 
we conducted a comparison between the estimated average and 5\%-tile percentile \ac{UE} throughput using both the ideal baseline and \ac{ABM}-friendly estimators, 
against the \ac{UE} throughput obtained from measurements in the test set. 
This assessment employed the formulation introduced in Section \ref{sec:rate-fia}. 
For this comparison, 
we employed the \ac{MAE} and \ac{MAPE} metrics. 
Additionally, we employed the mean estimator as a further benchmark, 
which characterizes the \ac{DL} \ac{UE} rate by computing the mean values of the related data observed in the training set.

\textcolor{black}{Regarding complexity, 
it is worth noting that our gradient boosting model required less than 1 minute per cell for training the ideal baseline or the ABM-friendly estimator in a machine powered by a Intel\textregistered $\:$ Core\textsuperscript{TM} i7-9700 CPU @ 3.00 GHz with 32~GB of Random Access Memory. The hyperparameters of the gradient boosting model were configured as follows: 
500 estimators, a maximum depth of 4, and a learning rate of 0.01.}

\begin{table}[]
\caption{Accuracy of the \ac{UE} rate model.}
\scriptsize
\centering
\begin{tabular}{l|cc|cc}
\hline
 & \multicolumn{2}{c|}{Mean \ac{UE} \ac{DL} Rate}&  \multicolumn{2}{c}{5\%-tile \ac{UE} \ac{DL} Rate} \\ 
{Algorithms}                            & ${\ac{MAE}}$          & ${\ac{MAPE}}$      & ${\ac{MAE}}$     & ${\ac{MAPE}}$        \\ \hline
\textit{Ideal baseline}                 & 2.24 Mbps             & 10.60\%            & 0.14 Mbps        & 27.94\%              \\ 
\textit{ABM-friendly}                   & 3.34 Mbps             & 15.74\%            & 0.14 Mbps        & 28.14\%              \\ 
\textit{Mean estimator}                 & 4.27 Mbps             & 21.52\%            & 0.17 Mbps        & 35.90\%              \\ \hline
\end{tabular}
\label{Table:UE DL Rate accuracy}
\end{table}

The results, 
presented in Table \ref{Table:UE DL Rate accuracy}, 
reveal that, as anticipated, 
the ideal baseline estimator exhibits the highest performance when evaluating the average and 5\%-tile percentile \ac{UE} throughput. 
Specifically, it achieves a \ac{MAE} of 2.24\,Mbps and 0.14\,Mbps, 
alongside a \ac{MAPE} of 10.60\,\% and 27.94\,\% for the average and 5\%-tile percentile \ac{UE} throughput, respectively. 
The 5\%-tile percentile \ac{UE} throughput, 
characterized by lower values than the average, 
incurs larger \ac{MAPE} values.

In contrast, 
the mean estimator demonstrates the weakest performance, 
yielding a \ac{MAE} of 4.27\,Mbps and 0.14\,Mbps, 
along with a \ac{MAPE} of 21.52\,\% and 35.9\,\%, for the average and 5\%-tile percentile \ac{UE} throughput, respectively. 
This represents a 2$\times$ and a 28\,\% increase in \ac{MAPE} when estimating the mean and the 5\%-tile \ac{UE} rate, respectively.

Significantly, 
the \ac{ABM}-friendly estimator strikes a commendable balance between complexity and performance. 
Specifically, it achieves a \ac{MAE} of 3.34\,Mbps and a \ac{MAPE} of 15.74\,\% for forecasting the average \ac{UE} rate,
and records a \ac{MAE} of 0.14,Mbps and a \ac{MAPE} of 28.14\,\% for predicting the 5\%-tile percentile \ac{UE} rate.
These figures represent a 48\,\% increase in \ac{MAPE} (1\,MBps in \ac{MAE}) compared to the ideal baseline estimator for the average \ac{UE} rate, 
and a negligible increase for the 5\%-tile \ac{UE} rate. 
This suggests that the inclusion of \ac{DL} traffic volume and the fraction of \ac{RSRP} samples below 100 dB offers limited accuracy improvement for this process at the cell-edge.
\section{Agent-Based Stochastic Carrier Shutdown Model}
\label{sec:abm_model}

Drawing inspiration from concepts in the Monte Carlo method \cite{Kroese2014WhyTM} and agent-based modeling \cite{Axelrod1997}, 
we introduce a novel, expert-based, white-box model in this section. 
This model replicates the behavior of a large-scale network when a carrier shutdown solution, 
as outlined in Section~\ref{sec:carriershutdown}, 
is implemented. 
With the aid of this model, 
we can extend the network's energy consumption and \ac{UE} rate predictions of the previous two sections to encompass any carrier shutdown and A4 handover parameter configurations.

As identified in such Sections \ref{sec:pc_model} and \ref{sec:rate_model}, 
it is essential to model the following parameters as functions of the carrier shutdown and A4 handover parameter setup in order to enable this generalization: 
\begin{itemize}
\item 
The number, $U^{\rm UE}_{i,h}$, of \ac{RRC} connected \acp{UE}.
\item 
The \ac{DL} \ac{PRB} load, $\frac{\Delta^{\rm DL}_{i,h}}{N_i^{\rm DL}}$.
\item 
The duration, $t^{\rm CS}_{i,h}$, of carrier shutdown for each cell, $i$, and hour, $h$.
\end{itemize}
With respect to the \ac{DL} \ac{PRB} load, $\frac{\Delta^{\rm DL}_{i,h}}{N_i^{\rm DL}}$,
and given that the number, $N_i^{\rm DL}$, of available \ac{DL} \acp{PRB} for each cell, $i$, is provided by the engineering parameters, 
our focus lies in modeling the number, $\Delta^{\rm DL}_{i,h}$, of utilized \ac{DL} \ac{PRB} for every cell, $i$, during each hour, $h$.

It is crucial to note that this expert-based, white-box modeling solely relies on unbiased information. 
As discussed in Section~\ref{sec:data},
this information is collected during an initial measurement campaign conducted prior to optimization. 
During this campaign, 
remind that all energy-saving solutions, 
including carrier shutdown, 
are deactivated to establish a baseline network energy efficiency and performance.

\subsection{Problem statement}
\label{sec:problemSetup}

Using the unbiased cell-level \acp{KPI} data set, \(d^{\rm KPI}\) (refer to Section \ref{sec:cell-level_KPIs}), 
and given the extensive data collected over multiple days, 
we can characterize the distributions, \(f_{U^{\rm UE}_{i,h}}\) and \(f_{\Delta^{\rm DL}_{i,h}}\), of the number, \(U^{\rm UE}_{i,h}\), of \ac{RRC} connected \acp{UE} and the number, \(\Delta^{\rm DL}_{i,h}\), of utilized \ac{DL} \acp{PRB} for each cell, \(i\), at each hour, \(h\), under the condition that the carrier shutdown solution is not activated. 
Our analysis suggests that the \acp{PDF}, \(f_{U^{\rm UE}_{i,h}}\) and \(f_{\Delta^{\rm DL}_{i,h}}\), can be approximated by Gaussian distributions.

\begin{definition}
    \label{def_model_input}
    \textbf{Inputs to our expert-based, white-box model}:
    We denote the set of stochastic inputs to our expert-based, white-box model as \({\cal X}^{\rm IN}_{i,h}\), 
    wherein this set encompasses, 
    among other elements, 
    the \acp{PDF}, \(f_{U^{\rm UE}_{i,h}} = {\cal N}(\mu_{U^{\rm UE}_{i,h}},\sigma_{U^{\rm UE}_{i,h}})\) and \(f_{\Delta^{\rm DL}_{i,h}} = {\cal N}(\mu_{\Delta^{\rm DL}_{i,h}}, \sigma_{\Delta^{\rm DL}_{i,h}})\), representing the number, \(U^{\rm UE}_{i,h}\), of \ac{RRC} connected \acp{UE} and the number, \(\Delta^{\rm DL}_{i,h}\), of utilized \ac{DL} \acp{PRB} for each cell, \(i\), at each hour, \(h\).
\end{definition}

It is important to note that our focus is on working days (Monday to Friday) and the typical 24-hour day. 
Additionally, due to the granularity of the cell-level \ac{KPI} data set, \(d^{\rm KPI}\), 
we have a single data sample per working day for each cell, \(i\), and typical hour, \(h\), to estimate such \acp{PDF}.
The \acp{PDF} are built using the many days of data. 

Given the aforementioned unbiased distributions, 
the network's engineering parameters (see Section \ref{sec:engineering_parameters}), 
and the carrier shutdown logic detailed in Section \ref{sec:carriershutdown}, 
our objective is to predict the network's behavior when the carrier shutdown solution is activated.

\begin{definition}
    \label{def_model_output}
    \textbf{Outputs of our expert-based, white-box model}:
    We denote the set of stochastic outputs of our expert-based, white-box model as \({\cal X}^{\rm OUT}_{i,h^\prime}\),
    wherein this set includes the predicted distributions, \(\hat{f}_{U^{\rm UE}_{i,h^\prime}}\), \(\hat{f}_{\Delta^{\rm DL}_{i,h^\prime}}\), and \(\hat{f}_{t^{\rm CS}_{i,h^\prime}}\), of the predicted number \(\hat{U}^{\rm UE}_{i,h^\prime}\) of \ac{RRC} connected \acp{UE}, the predicted number, \(\hat{\Delta}^{\rm DL}_{i,h^\prime}\), of utilized \ac{DL} \acp{PRB}, and the predicted duration, \(\hat{t}^{\rm CS}_{i,h^\prime}\), of carrier shutdown, for each cell, \(i\), and hour, \(h^\prime\), respectively.
\end{definition}

\subsection{Dealing with stochasticity - Monte Carlo method}
\label{sec:MCmethod}

To account for the stochastic nature of traffic and channel conditions, 
and to forecast performance at each hour, \(h^\prime\), 
we employed the Monte Carlo method \cite{Kroese2014WhyTM}.

Our Monte Carlo method encompasses \(R\) runs or snapshots. 
Each individual Monte Carlo run, 
denoted as \(\texttt{r}\), 
is initialized with data drawn from the unbiased distributions, \(f_{U^{\rm UE}_{i,h}}\) and \(f_{\Delta^{\rm DL}_{i,h}}\) within \({\cal X}^{\rm IN}_{i,h}\) (as defined in Definition \ref{def_model_input}).

\begin{definition}
    \label{def_MC_input}
    \textbf{Inputs to each Monte Carlo run, \(\texttt{r}\)}:
    Let \({\cal X}^{\rm MC-IN}_{i,h^\prime,r}\) represent the set of inputs utilized to seed each Monte Carlo run, \(\texttt{r}\), 
    where this set includes the random realizations, \(\tilde{U}^{\rm UE}_{i,h^\prime,r}\) and \(\tilde{\Delta}^{\rm DL}_{i,h^\prime,r}\), drawn from the unbiased distributions, \(f_{U^{\rm UE}_{i,h}}\) and \(f_{\Delta^{\rm DL}_{i,h}}\) within \({\cal X}^{\rm IN}_{i,h}\).
\end{definition}

The anticipated network behavior at each hour, \(h^\prime\), 
namely \(\hat{f}_{U^{\rm UE}_{i,h^\prime}}\), \(\hat{f}_{\Delta^{\rm DL}_{i,h^\prime}}\), and \(\hat{f}_{t^{\rm CS}_{i,h^\prime}}\) within \({\cal X}^{\rm OUT}_{i,h^\prime}\) 
(as defined in Definition \ref{def_model_output}), 
is subsequently derived from statistics obtained across all Monte Carlo runs \(\texttt{r} \in \{1, \cdots, \texttt{R}\}\),
as it will be discussed later in this section.

\begin{definition}
    \label{def_MC_output}
    \textbf{Outputs of each Monte Carlo run, \(\texttt{r}\)}:
    Let \({\cal X}^{\rm MC-OUT}_{i,h^\prime,r}\) symbolize the set of outputs for each Monte Carlo run, \(\texttt{r}\), 
    wherein this set encompasses the predicted number, \(\hat{U}^{\rm UE}_{i,h^\prime,r}\), of \ac{RRC} connected \acp{UE}, the predicted number, \(\hat{\Delta}^{\rm DL}_{i,h^\prime,r}\), of used \ac{DL} \acp{PRB}, and the predicted duration, \(\hat{t}^{\rm CS}_{i,h^\prime,r}\), of the carrier shutdown for each cell, \(i\), hour, \(h^\prime\), and Monte Carlo run, \(\texttt{r}\).
\end{definition}


It is worth noting that our Monte Carlo method incorporates a rolling concept from hour to hour. 
This implies that not only the unbiased data distributions but also the predicted statistics at hour, \(h^\prime\), if available, serve as inputs to our expert-based, white-box model to predict network behavior at the subsequent hour \(h^\prime +1\). 
This concept will also be elaborated on further in this section.

\subsection{Dealing with carrier shutdown - Agent-based modelling}
\label{sec:ABMmethod}

To model the outputs, \({\cal X}^{\rm MC-OUT}_{i,h^\prime,r}\), of a given Monte Carlo run, \(\texttt{r}\), in relation to any carrier shutdown and A4 handover parameter configuration, 
we have developed a customized Agent-Based Model (\ac{ABM}).

\acp{ABM} have found extensive application in economics for simulating macroeconomic structures emerging from the repeated local interactions among socioeconomic agents.

The fundamental concept involves constructing a virtual environment and populating it with agents, 
each endowed with distinct attributes. 
These agents adhere to fundamental guidelines governing their interactions with both each other and their surroundings. 
Typically, these guidelines are grounded in insights about behavior and the local environment. 
Consequently, \acp{ABM} are dynamic, stochastic systems, usually executed on computers, 
evolving over time through iterative processes or algorithms. 
During these processes, 
agents are adjusted based on established rules. 
Often, \acp{ABM} incorporate randomness, 
where agents select various behavioral options randomly. 
Consequently, Markov chain theory is well-suited for the mathematical formalization of \acp{ABM} \cite{Banisch2015}.

In this context, 
we propose an \ac{ABM} defined by a set of \(C\) agents, 
one for each capacity cell, \(c \in {\cal C}_C\), 
with each agent characterized by individual attributes drawn from a finite list of possibilities
---specifically, \(\texttt{active}\) or \(\texttt{shutdown}\). 
We denote the set of possible attributes as \({\cal S} = \{\texttt{active} = 1, \texttt{shutdown} = 0\}\), 
referring to the solution space as \(\Sigma\) and an agent configuration as \(x \in \Sigma\), 
with \(x = (x_1, \cdots, x_c, \cdots, x_C)\). 
Consequently, the cardinality of the solution space is \(2^\Sigma\). 
The process of updating agent attributes at each time step, \(\texttt{t}\), comprises two parts. 
First, a random subset of agents is chosen based on a probability distribution, \(\omega\). 
Then, the agents' attributes are updated according to a rule, \(u\), determined by the subset of agents chosen at that time. 
Both of these processes will be detailed in Subsections \ref{sec:update_rule} and \ref{sec:agent_selection}.

It is important to note that, 
due to our implementation via a sequential update scheme, 
only one agent 
---equivalent to one capacity cell, $c \in {\cal C}_C$--- 
can alter its attribute at a given time. 
This signifies that transitions can only occur between agent configurations that vary in at most one bit, 
following a bit representation.  
The agent selection and updating processes are sequentially iterated for a set number of time steps until a stable agent configuration, \(x\), is reached. 
An agent configuration, \(x\), is considered stable if no agent can change its attribute within a time step, \(\texttt{t}\),
i.e. no cell can be shutdown or activated anymore at this hour. 
To prevent ping-pong effects or loops, 
the maximum number of time steps, \(\texttt{t}\), within a Monte Carlo run, \(\texttt{r}\), is constrained by \(\texttt{t}_{\texttt{max}}\). 
With this specification, 
a Monte Carlo run, \(\texttt{r}\), can be perceived as a random walk across the solution space, \(\Sigma\). 
It is important to note that, 
at the start of the first run in the initial hour, i.e. \(h^\prime = 1\), \(\texttt{r} = 1\), and \(\texttt{t} = 1\), 
we assume that all capacity cells are active, i.e. \(x_c = 1\) for all \(c \in {\cal C}_C\).

\subsubsection{Updating rule check}
\label{sec:update_rule}

To expedite the processing of the \ac{ABM}, 
we implemented a preliminary updating rule check.

Following the principles outlined in Section \ref{sec:sd_logic}, 
we define the set, \({\cal C}^{\rm \rightarrow sd}\), as the set of capacity cells active at hour, \(h^\prime\), Monte Carlo run, \(\texttt{r}\), and time step, \(\texttt{t}\), that can alter their attribute 
---specifically, cells that can be shut down. 
This set is identified based on a rule, \(u_{\rm sd}\), comprising three conditions:

\begin{enumerate}
    \item 
    $\tilde{U}^{\rm UE}_{c,h^\prime,r} <  \chi^{\rm UE}_{c,h^\prime}$, and
    \item 
    $\tilde{\Delta}^{\rm DL}_{c,b(c),h^\prime,r} = \tilde{\Delta}^{\rm DL}_{c,h^\prime,r} + \tilde{\Delta}^{\rm DL}_{b(c),h^\prime,r} < \chi^{\rm DL}_{c,b(c),h^\prime}$, and 
    \item 
    $\tilde{\Delta}^{\rm UL}_{c,b(c),h^\prime,r} = \tilde{\Delta}^{\rm UL}_{c,h^\prime,r} + \tilde{\Delta}^{\rm UL}_{b(c),h^\prime,r} < \chi^{\rm UL}_{c,b(c),h^\prime}$.
\end{enumerate}

In a similar manner, 
we identify the set, \({\cal C}^{\rm \rightarrow ac}\), as the set of capacity cells that are shut down at hour, \(h^\prime\), Monte Carlo run, \(\texttt{r}\), and time step, \(\texttt{t}\), and have the potential to change their attribute
---these cells can be reactivated. 
This set is determined using a rule, \(u_{\rm ac}\), comprising three conditions:

\begin{enumerate}
    \item 
    $U^{\rm UE}_{b(c),h^\prime,r} > \Psi^{\rm UE}_{c,b(c),h^\prime}$, or
    \item 
    $\Delta^{\rm DL}_{b(c),h^\prime,r} > \Psi^{\rm DL}_{c,b(c),h^\prime}$, or 
    \item 
    $\Delta^{\rm UL}_{b(c),h^\prime,r} > \Psi^{\rm UL}_{c,b(c),h^\prime}$.
\end{enumerate}

If either the updating rule, \(u_{\rm sd}\) or \(u_{\rm ac}\), is satisfied, 
specific capacity cells can either initiate the shutdown process or the reactivation process, respectively, 
depending on their state.

If no capacity cell meets the conditions of either updating rule, \(u_{\rm sd}\) or \(u_{\rm ac}\), 
meaning, \({\cal C}^{\rm \rightarrow sd}  \cup  {\cal C}^{\rm \rightarrow sd} = \emptyset\), 
the current agent configuration, \(x\), becomes stable, 
and the Monte Carlo run, \(\texttt{r}\), terminates. 
The \ac{ABM} then proceeds to the next run or advances to the next hour 
(if the maximum number of runs, \(\texttt{R}\), has been reached), 
or concludes (if it was already processing the last hour).

\subsubsection{Agent selection}
\label{sec:agent_selection}

The agent selection at each time step, \(\texttt{t}\), of Monte Carlo run, \(\texttt{r}\), at hour, \(h^\prime\), is founded on our novel concept of ``load distance to threshold". 
This concept is grounded in the following logic: 
Among the capacity cells in the set, \({\cal C}^{\rm \rightarrow sd}\) (cells that are active and can be shut down),
and considering a given shutdown entry condition \ac{DL} \ac{PRB} threshold, \(\chi^{\rm DL}\), 
the capacity cell, \(c\), with the smallest combined load ---combining its load and that of its paired coverage cell, \(\Delta^{\rm DL}_{c,b(c)}\)--- is more likely to be shut down first or more frequently compared to other cells. 
Here, \(b(c)\) refers to the paired coverage cell of capacity cell, \(c\). 
Similarly, within the set, \({\cal C}^{\rm \rightarrow ac}\) (cells that are shut down and can reactivate), 
and considering a shutdown leaving condition \ac{DL} \ac{PRB} threshold, \(\Psi^{\rm DL}\), 
the capacity cell, \(c\), with the highest load on its paired coverage cell, \(\Delta^{\rm DL}_{b(c)}\), is more likely to reactivate first or more frequently compared to other cells. 
Additional details on the shutdown entry and leaving conditions can be found in Section \ref{sec:sd_logic}.

Building on these observations,
we begin by computing the ``load distance" in terms of the number of \ac{DL} \acp{PRB} to the shutdown entry condition \ac{DL} \ac{PRB} threshold for each capacity cell, $c$,  in the set, \({\cal C}^{\rm \rightarrow sd}\), denoted as \(d_{c,h^\prime,r} = \chi^{\rm DL}_{c,b(c)} - \tilde{\Delta}^{\rm DL}_{c,b(c),h^\prime,r}\). 
Similarly, for the set, \({\cal C}^{\rm \rightarrow ac}\), we compute the "load distance" to the shutdown leaving condition \ac{DL} \ac{PRB} threshold as \(d_{c,h^\prime,r} = \tilde{\Delta}^{\rm DL}_{b(c),h^\prime,r} - \Psi^{\rm DL}_{c,b(c)}\). 
Subsequently, we normalize each of these distances by dividing it by the sum of distances across all cells in both sets, 
resulting in \(d^{\rm N}_{c,h^\prime,r} = \frac{d_{c,h^\prime,r}}{\sum_{{\cal C}^{\rm \rightarrow sd} \cup {\cal C}^{\rm \rightarrow sd}} d_{c,h^\prime,r}}\). 
The engineering parameters provide the thresholds, \(\chi^{\rm DL}_{c,b(c)}\) and \(\Psi^{\rm DL}_{c,b(c)}\), 
while the quantities of used \ac{DL} \acp{PRB}, \(\tilde{\Delta}^{\rm DL}_{c,b(c),h^\prime,r}\) and \(\tilde{\Delta}^{\rm DL}_{b(c),h^\prime,r}\), are initially drawn from the unbiased distributions, \(f_{\Delta^{\rm DL}_{i,h}}\), of the corresponding cells, \(c\) and \(b(c)\), at the beginning of Monte Carlo run, \(\texttt{r}\) (as per Definition \ref{def_MC_input}). 
These quantities are then updated as described in Section \ref{sec:updating} when \(\texttt{t} > 1\).

Using these normalized estimations, \(d^{\rm N}_{c,h^\prime,r}\) for all \(c \in {\cal C}^{\rm \rightarrow sd} \cup {\cal C}^{\rm \rightarrow sd}\), 
we construct the \ac{PMF}, \(\omega\), 
and select a capacity cell, \(c^{\rm a}\), from this distribution. 
If the chosen capacity cell, \(c^{\rm a}\), is active, 
it initiates the shutdown process,
and assesses whether its connected \acp{UE}, \({\cal U}_{c^{\rm a}} = \{1,\cdots,u,\cdots, U_{c^{\rm a}}\}\), can be handed over to neighboring cells through an A4 inter-frequency handover.
Conversely, 
if the selected capacity cell, \(c^{\rm a}\), is shut down, 
its paired coverage cell, \(b(c^{\rm a})\), starts the reactivation process. 
The \acp{UE} originally connected to this capacity cell, \(c^{\rm a}\),
as determined by the unbiased input data drawn at the beginning of Monte Carlo run, \(\texttt{r}\) (as defined in \ref{def_MC_input}),
are returned to it. 
Additional information regarding \ac{UE} handovers is available in Subsection \ref{sec:user_transfer}.

The distance to threshold metric that we have designed effectively implements the earlier-discussed logic, 
as it ensures that the capacity cell with the greatest distance to threshold is most likely to be selected and consequently take action first. 
The stochastic nature of this process should also be noted.

\subsubsection{User transfer}
\label{sec:user_transfer}

When capacity cell, $c^{\rm a}$, starts its shutdown process,
it commands its set of connected \acp{UE}, ${\cal U}_{c^{\rm a}}$, to perform inter-frequency handovers to the frequency of its paired coverage cell, $b(c^{\rm a})$, through the A4 handover entry condition (see Section \ref{sec:ue_transfer_logic}).
We denote by ${\cal I}_{c^{\rm a}}$ the set of cells in such frequency.

In this work, 
we use the stochastic model presented in \cite{DeDomenico2023} to estimate the probability, $p^{\rm HO}_{c^{\rm a},j,h^\prime}$, of one such \ac{UE} being handed by capacity cell, $c^{\rm a}$, over a neighboring cell, $j \in {\cal I}_{c^{\rm a}}$, at hour, $h^\prime$.
Note the algorithm presented in \cite{DeDomenico2023} makes use of the engineering parameters, $d^{\rm EP}$, and the \ac{MR}, $d^{\rm MR}$, data sets (see Section \ref{sec:data}). 

\begin{definition}
    \label{def_transferProb}
    {\textbf{Transfer probability, $p^{\rm HO}_{c^{\rm a},j,h^\prime}$}}:
    Let us denote by $p^{\rm HO}_{c^{\rm a},j,h^\prime}$ the probability of a \ac{UE} transferring from capacity cell, $c^{\rm a}$, to a neighboring cell, $j \in {\cal I}_{c^{\rm a}}$, when performing an inter-frequency handover through the A4 handover entry condition, at hour, $h^\prime$. 
\end{definition}


To decide to which specific neighbouring cell, $j \in {\cal I}_{c^{\rm a}}$, each \ac{UE}, $u$, of capacity cell, $c_{a}$, is handed over at hour, $h^\prime$,
we build an empirical \ac{PMF} using all values, $p^{\rm HO}_{c^{\rm a},j,h^\prime}$, from all neighbouring cells, $j \in {\cal I}_{c^{\rm a}}$,
and draw a target neighbouring cell, $j^{\rm t}$, from it. 

Importantly, if \acp{UE} cannot be handed over any neighbouring cell due to, e.g. the existence of coverage holes
---no suitable neighbour exits---,
i.e.  $p^{\rm HO}_{c^{\rm a},j,h^\prime}=0, \forall j \in {\cal I}_{c^{\rm a}}$,
the capacity cell, $c^{\rm a}$, aborts its shutdown process,
and the sequential process of the Monte Carlo run moves to the next time step, $\texttt t + 1$. 

\subsubsection{Updating and storing of Monte Carlo run statistics}
\label{sec:updating}

At this juncture, 
the capacity cell, $c^{\rm a}$, can be in one of two operational states: 
either it is in the process of shutting down or it is undergoing reactivation.

In the former scenario, 
if all \acp{UE}, ${\cal U}_{c^{\rm a}}$, within the capacity cell, $c^{\rm a}$, can be seamlessly handed over to a neighboring cell, $j \in {\cal I}_{c^{\rm a}}$, 
meaning that the sum of the related handover probability is larger than zero, i.e. $\sum_j p^{\rm HO}_{c^{\rm a},j,h^\prime}>0$, 
the capacity cell, $c^{\rm a}$, proceeds to shut down. 
As a consequence, 
the agent configuration vector, $x = (x_1,\cdots,x_c,\cdots,x_C)$, is updated, 
specifically flipping the value of $x_{c^{\rm a}}$ from 1 to 0. 
In addition, the set of outputs denoted as ${\cal X}^{\rm MC-OUT}_{i,h^\prime,r}$ (as described in Definition \ref{def_MC_output}) is revised as follows:
For each \ac{UE}, $u$, within set, ${\cal U}_{c^{\rm a}}$, that transfers to neighboring cell, $j^{\rm t}$, 
the predicted number, $\hat{U}^{\rm UE}_{j^{\rm t},h^\prime,r}$, of \ac{RRC} connected \acp{UE} in cell, $j^{\rm t}$, for that particular hour, $h^\prime$, and Monte Carlo run, $\texttt r$, is incremented by one unit. 
Simultaneously, the counter associated with the capacity cell, $c^{\rm a}$, is set to 0, i.e. $\hat{U}^{\rm UE}_{c^{\rm a},h^\prime,r} = 0$.
Furthermore, the projected number, $\hat{\Delta}^{\rm DL}_{j^{\rm t},h^\prime,r}$, of utilized \ac{DL} \acp{PRB} in cell, $j^{\rm t}$, at the specified hour, $h^\prime$, and Monte Carlo run, $\texttt r$, is augmented by $\frac{\hat{\Delta}^{\rm DL}_{c^{\rm a},h^\prime,r}}{U_{c^{\rm a}}}$, 
with the assumption that all \acp{UE}, $u$, within set, ${\cal U}_{c^{\rm a}}$, on average employ the same number of \acp{PRB}. 
Similarly, the counter tied to the capacity cell, $c^{\rm a}$, is set  to 0, i.e. $\hat{\Delta}^{\rm DL}_{c^{\rm a},h^\prime,r} = 0$. 
In line with the granularity of 1 hour that is being employed, 
the anticipated duration, $\hat{t}^{\rm CS}_{i,h^\prime,r}$, of the carrier shutdown for capacity cell, $c^{\rm a}$, during the specified hour, $h^\prime$, and Monte Carlo run, $\texttt r$, is set to 60 minutes.

Importantly, 
our \ac{ABM} stores all these movements of \acp{UE} and \acp{PRB} from the capacity cell, $c^{\rm a}$, to each of its neighbouring cells, $j \in {\cal I}_{c^{\rm a}}$, at hour, $h^\prime$, Monte Carlo run, $\texttt r$, and time step, $\texttt t$, in a replay memory where all relevant actions are recorded. 

In the latter case, 
if the capacity cell, $c^{\rm a}$, is waking up, 
the agent configuration, $x = (x_1,\cdots,x_c,\cdots,x_C)$, is modified,
with $x_{c^{\rm a}}$ flipping from 0 to 1.
We then retrieve, 
from the previously mentioned replay memory, 
the \acp{UE} connected to ---and the used \ac{DL} \acp{PRB} in--- capacity cell, $c^{\rm a}$, at the beginning of this Monte Carlo run, $\texttt r$, 
i.e. at $\texttt t=1$, at this hour, $h^\prime$, 
and hand them back to it, 
updating the set of outputs, ${\cal X}^{\rm MC-OUT}_{i,h^\prime,r}$, accordingly. 
Note that the replay memory is particularly useful to identify in which neighbouring cells, 
the original \acp{UE} of capacity cell, $c^{\rm a}$, are, 
and subtract them and their numbers of \acp{PRB} from their statistics.  
Note that in this case, the predicted duration, $\hat{t}^{\rm CS}_{i,h^\prime,r}$, of the carrier shutdown of capacity cell, $c^{\rm a}$, at hour, $h^\prime$, and Monte Carlo run, $\texttt r$, is set to 0 minutes.

\subsubsection{Termination and expert-based, white-box model statistics}
\label{sec:termination}

Once all the output variables within the set, ${\cal X}^{\rm MC-OUT}_{i,h^\prime,r}$, have been updated for a particular time step, $\texttt t$, 
the \ac{ABM} proceeds to the subsequent time step, $\texttt t+1$. 
At this juncture, 
a fresh evaluation of updating rules, agent selection, \ac{UE} transfers, and the subsequent update and storage of statistics is carried out, 
as delineated in the preceding subsections. 
This iterative progression persists until the current Monte Carlo run, $\texttt r$, attains stability, 
meaning no agent can modify its attributes due to the inability of any candidate to satisfy the updating rules, 
or until the maximum allowed number of time steps, $\texttt t_{\texttt{max}}$, within a Monte Carlo run is reached.

Upon reaching either of these conditions, 
the \ac{ABM} proceeds to the subsequent Monte Carlo run, $\texttt r+1$, 
at which point the entire process described above is replicated. 
However, in this instance, 
a fresh set of random input variables, ${\cal X}^{\rm MC-IN}_{i,h^\prime,r}$ (as defined in Definition \ref{def_MC_input}), is drawn to initialize this new Monte Carlo run.

In the event that the maximum specified number of runs, $\texttt R$, is reached, 
the \ac{ABM} moves on to the next hour, $h^\prime+1$, 
unless it is already processing the final hour. 
If this is the case, 
the simulation concludes.

A crucial point to emphasize is that, 
at the conclusion of each hour, $h^\prime$, 
the \ac{ABM} computes the set, ${\cal X}^{\rm OUT}_{i,h^\prime}$, of stochastic outputs generated by our expert-based, white-box model for that hour. 
This calculation is performed using the set, ${\cal X}^{\rm MC-OUT}_{i,h^\prime,r}$, encompassing the outputs from each Monte Carlo run, $\texttt r \in \{1,\cdots, \texttt R\}$. For example, 
the anticipated average duration, $\hat{t}^{\rm CS}_{i,h^\prime}$, of carrier shutdowns for each capacity cell, $c$, during hour, $h^\prime$, can be computed as $\hat{t}^{\rm CS}_{i,h^\prime} = \frac{\sum_{\{1,\cdots, \texttt R\}}\hat{t}^{\rm CS}_{i,h^\prime,r}}{\texttt R}$. 
This same reasoning can be applied to other variables as well. 
Importantly, by utilizing the results from the $\texttt R$ Monte Carlo runs for the hour, 
the \ac{ABM} is capable of providing not only average statistics but also distributions, 
as previously discussed.

\subsection{Rolling from hour to hour}

At the onset of the first Monte Carlo run of the initial hour, 
specifically when $h^\prime=1$, $\texttt r=1$, and $\texttt t=1$, 
we made the assumption that all capacity cells are in an active state, 
i.e $x_c = 1$, $c \in {\cal C}_C$. 
Consequently, 
by the end of this first hour, 
we achieve $\texttt R$ stable Monte Carlo runs each with a corresponding stable agent configuration, 
that is, $x_{h^\prime=1,r} = (x_1,\cdots,x_c,\cdots,x_C)$ for all $\texttt r \in \{1,\cdots, \texttt R \}$.

Upon transitioning from the present hour, $h^\prime$, to the subsequent hour, $h^\prime + 1$, 
two scenarios may unfold: 
\emph{i)}  All capacity cells are activated anew. 
This scenario allows for the independent processing of different hours.
\emph{ii)}  A subset of capacity cells may be deactivated based on outcomes from the prior hour, $h^\prime$. 
In this case, 
the hours are no longer independent, 
and a rolling process must be implemented.

In our implemented approach, 
during the transition from one hour, $h^\prime$, to the next, $h^\prime + 1$, 
a starting agent configuration for the new hour is randomly selected from the replay memory. 
This selection is made among the $\texttt R$ stable agent configurations from the preceding hour, $h^\prime$. 
It is worth noting that the set of inputs, ${\cal X}^{\rm MC-IN}_{i,h^\prime+1,r}$, in the subsequent hour, $h^\prime + 1$, differs from that in the previous hour, $h^\prime$, due to unbiased traffic statistics evolving from hour to hour, 
i.e. $f_{U^{\rm UE}_{i,h}} \neq f_{U^{\rm UE}_{i,h+1}}$ and $f_{\Delta^{\rm DL}_{i,h}} \neq f_{\Delta^{\rm DL}_{i,h+1}}$. 
Consequently, 
to calculate the statistics for the chosen starting agent configuration, 
a simulation of the carrier shutdown process from a state with all capacity cells active to the selected starting agent configuration is needed. 
This simulation involves the requisite \ac{UE} transfers as well as updates and storage of relevant statistics. 
Importantly, this simulation does not necessitate a review of updating rules or a stochastic agent selection, 
as the desired agent configuration is already known. 
Once the chosen agent configuration is attained, 
the Monte Carlo run can be initiated.

\bigskip

Algorithm \ref{alg:ABM} summarises this fused Monte Carlo and \ac{ABM} framework. 

\begin{algorithm*}
\scriptsize
  \SetKwData{Left}{left}
  \SetKwData{This}{this}
  \SetKwData{Up}{up}
  \SetKwFunction{Union}{Union}
  \SetKwFunction{FindCompress}{FindCompress}
  \SetKwInOut{Input}{input}
  \SetKwInOut{Output}{output}
  \SetKwData{And}{and}
  \SetKwData{Break}{break}
  \SetKwData{Continue}{continue}
  \caption{Carrier shutdown, expert-based, white-box model  \label{alg:ABM} }
  \Input{Engineering parameters data sets, $d^{\rm EP}$,
  Measurement report data set, $d^{\rm MR}$, and 
  Unbiased distributions, $f_{U^{\rm UE}_{i,h}}$ and $f_{\Delta^{\rm DL}_{i,h}}$ in ${\cal X}^{\rm IN}_{i,h}$ $\forall i \in {\cal C}, h \in \{1,\cdots, h,\cdots, 24\} $ (see Definition \ref{def_model_input}), 
  extracted from cell-level \acp{KPI} data set, $d^{\rm KPI}$}
  \Output{Predicted distributions, $\hat{f}_{U^{\rm UE}_{i,h^\prime}}$, $\hat{f}_{\Delta^{\rm DL}_{i,h^\prime}}$ and $\hat{f}_{t^{\rm CS}_{i,h^\prime}}$ in ${\cal X}^{\rm OUT}_{i,h^\prime}$ $\forall i \in {\cal C}, h^\prime \in \{1,\cdots, h^\prime,\cdots, 24\} $ (see Definition \ref{def_model_output})}
  \BlankLine
  \For{$h^\prime \in \{1,\cdots, h^\prime,\cdots, 24\}$}{ 
    \For{$\texttt r \in \{1,\cdots, \texttt r,\cdots, \texttt R\}$}{
         
        ${\cal X}^{\rm MC-IN}_{i,h^\prime,r}  \leftarrow \tilde{U}^{\rm UE}_{i,h^\prime,r}  \leftarrow  f_{U^{\rm UE}_{i,h}}$ \And
        ${\cal X}^{\rm MC-IN}_{i,h^\prime,r}  \leftarrow \tilde{\Delta}^{\rm DL}_{i,h^\prime,r}  \leftarrow  f_{\Delta^{\rm DL}_{i,h}}$         
        \tcp{Draw the set of inputs, ${\cal X}^{\rm MC-IN}_{i,h^\prime,r}$, of this Monte Carlo run (see Definition\ref{def_MC_input}) } 

        \eIf{$h^\prime == 1$ \And $\texttt r == 1$ }{
            $x_c = 1, \forall c \in {\cal C}_C$ 
            \tcp{Initialize the agent configuration, $x = \{1,\cdots, x_c,\cdots, X_C\} $ }      
        }{
            \tcc{Use the rolling process from hour to hour to get the starting agent configuration}   
        }
        
        \tcc{Start random walk within the Monte Carlo run}
        \For{$\texttt t \in \{1,\cdots, \texttt t,\cdots, \texttt t_{\texttt{max}}\}$ }{

            \tcc{Updating rule check}
            \For{$c \in {\cal C}_{C} $}{
                \eIf{$x_c == 1$ \And $u_{\rm sd} == 1$}{
                    ${\cal C}^{\rm \rightarrow sd} \leftarrow c$ 
                    \tcp{Identify the capacity cells that can be shut down (rule, $u_{\rm sd}$)}
                } {
                    ${\cal C}^{\rm \rightarrow ac} \leftarrow c$ 
                    \tcp{Identify the capacity cells that can be reactivated (rule, $u_{\rm ac}$)}
                }
            }

            \tcc{Check if a stable agent configuration is found --- Termination of Monte Carlo run}
            \If{${\cal C}^{\rm \rightarrow sd}  \cup  {\cal C}^{\rm \rightarrow sd}$}{    
                \Break \;
            } 
            
            \tcc{Agent selection }
            \For{$c \in {\cal C}^{\rm \rightarrow sd} $}{
                $d_{c,h^\prime,r} = \chi^{\rm DL}_{c,b(c)} - \tilde{\Delta}^{\rm DL}_{c,b(c),h^\prime,r}$ 
                \tcp{Compute distance to threshold for cells in set, ${\cal C}^{\rm \rightarrow sd}$}
            } 
            \For{$c \in {\cal C}^{\rm \rightarrow ac} $}{
                $d_{c,h^\prime,r} = \tilde{\Delta}^{\rm DL}_{b(c),h^\prime,r} - \Psi^{\rm DL}_{c,b(c)}$ 
                \tcp{Compute distance to threshold for cells in set, ${\cal C}^{\rm \rightarrow ac}$}
            }
            $d^{\rm N}_{c,h^\prime,r} = \frac{d_{c,h^\prime,r}}{\sum_{{\cal C}^{\rm \rightarrow sd}  \cup  {\cal C}^{\rm \rightarrow sd}} d_{c,h^\prime,r}}$ 
            \tcp{Normalise distance to threshold}
            $\omega = {\rm PMF}(d^{\rm N}_{c,h^\prime,r}) \forall c \in {\cal C}^{\rm \rightarrow sd}  \cup  {\cal C}^{\rm \rightarrow sd} $
            \tcp{Create \ac{PMF}, from which a capacity cell will be selected at random to change its attribute}
            $c^{\rm a} \leftarrow \omega$ 
            \tcp{Select at random the capacity cell to change its attribute}

            \tcc{User transfer }
            \eIf{$x_{c^{\rm a}} == 1$}{
                \emph{Derive ${\cal I}_{c^{\rm a}}$ and calculate $p^{\rm HO}_{c^{\rm a},j,h^\prime} \forall j \in {\cal I}_{c^{\rm a}}$}
                \tcp{Derive the statistics necessary to drive the UE transfer of the selected capacity cell, 
                using information of data sets, $d^{\rm EP}$ and $d^{\rm MR}$ }
                \eIf{$p^{\rm HO}_{c^{\rm a},j,h^\prime} == 0 \forall j \in {\cal I}_{c^{\rm a}}$}{
                    \Continue
                    \tcp{If a UE cannot be transferred, the shutdown process is abandoned}
                } {
                    
                    $x_{c^{\rm a}} = 0$ 
                    \tcp{Shut down the selected capacity cell}
                    
                    \tcc{Updating and storing of statistics}
                    $\hat{U}^{\rm UE}_{c^{\rm a},h^\prime,r} = 0$ \And 
                    $\hat{\Delta}^{\rm DL}_{c^{\rm a},h^\prime,r} = 0$ \And
                    $\hat{t}^{\rm CS}_{c^{\rm a},h^\prime,r} = 60$\,min\;
                    \For{$u \in {\cal U}_{c^{\rm a}} $}{  
                        $j^{\rm t} \leftarrow {\rm PMF}(p^{\rm HO}_{c^{\rm a},j,h^\prime}) \forall j \in {\cal I}_{c^{\rm a}} $ \;
                        $\hat{U}^{\rm UE}_{j^{\rm t},h^\prime,r} = \hat{U}^{\rm UE}_{j^{\rm t},h^\prime,r} + 1$ \And
                        $\hat{\Delta}^{\rm DL}_{j^{\rm t},h^\prime,r} = \hat{\Delta}^{\rm DL}_{j^{\rm t},h^\prime,r} + \frac{\hat{\Delta}^{\rm DL}_{c^{\rm a},h^\prime,r}}{U_{c^{\rm a}}}$ \;   
                        }
                }
            }{
                
                $x_{c^{\rm a}} = 1$ 
                \tcp{Reactivate the selected capacity cell}
                
                \tcc{Updating and storing of statistics} 
                $\hat{U}^{\rm UE}_{c^{\rm a},h^\prime,r} = U^{\rm UE}_{c^{\rm a},h^\prime,r} $ \And
                $\hat{\Delta}^{\rm DL}_{c^{\rm a},h^\prime,r} = U^{\rm UE}_{c^{\rm a},h^\prime,r} $ \And
                $\hat{t}^{\rm CS}_{c^{\rm a},h^\prime,r} = 0$\,min\;
                \emph{In addition, discount number of \acp{UE} and used \ac{DL} \acp{PRB} from the respective counters of the respective cells hosting the \acp{UE} of the awaking capacity cell, $c^{\rm a}$}\;   
            } 
            \tcc{Updating of replay memory} 
            \emph{Save the agent configuration and related statistic of this hour, $h^\prime$, Monte Carlo run, $\texttt r$, and time step, $\texttt t$, in the replay memory  } 
        }  
        
        $ {\cal X}^{\rm MC-OUT}_{i,h^\prime,r}  \leftarrow  \hat{U}^{\rm UE}_{i,h^\prime,r} , \hat{\Delta}^{\rm DL}_{i,h^\prime,r}$ and $\hat{t}^{\rm CS}_{i,h^\prime, r} \forall i \in {\cal C}$ 
        \tcp{Update set, ${\cal X}^{\rm MC-OUT}_{i,h^\prime,r}$, of this Monte Carlo run (see Definition \ref{def_MC_output}). Note that the last entry to the replay memory belongs to the stable agent configuration and statistics of this Monte Carlo run, $\texttt r$, at this hour, $h^\prime$}
 
    }
    
    ${\cal X}^{\rm OUT}_{i,h^\prime}  \leftarrow  {\cal X}^{\rm MC-OUT}_{i,h^\prime,r} \forall i \in {\cal C} $       
    \tcp{Obtain the set, ${\cal X}^{\rm OUT}_{i,h^\prime}$, of outputs of our expert-based, white-box model for this hour, $h^\prime$, from the set of outputs, ${\cal X}^{\rm MC-OUT}_{i,h^\prime,r}$, of its $\texttt R$ Monte Carlo runs } 
  }
  \label{algo_disjdecomp}
\end{algorithm*}

\section{Experiments and Analysis}
\label{sec:results}

In this section, we assess the performance of the proposed energy-saving modeling framework in estimating network performance during carrier shutdown activation.

To evaluate the framework introduced in Section \ref{sec:abm_model}, we employ datasets from a real network in a metropolitan area of China. Our focus is on a set denoted as $\mathcal{C}$, comprising 657 cells categorized into (i) 375 capacity booster cells (constituting the set ${\cal C_{\rm C}}$) and (ii) 282 coverage cells (comprising the set ${\cal C_{\rm B}}$). These datasets spanned 12 days in April 2022 and were crucial for constructing the models detailed in Sections \ref{sec:pc_model}, \ref{sec:rate_model}, \ref{sec:abm_model}, and \ref{sec:benchmark}. Additionally, data from a 3-day period in May 2023 were utilized to create the test set and calculate the 'ground truth', vital for validating our model results.
{\color {black}Ground truth denotes the actual measured truth, 
as measured directly from our dataset.}
It is important to note that each dataset, and thus our ground truth, has been aggregated on an hourly basis, providing 24 values per day for each cell. These values offer insights into traffic, energy consumption, and measurement report statistics (see Table \ref{tab:Cell-level_KPIs} and \ref{tab:UE_measurement_reports}).

We benchmark our framework's performance with respect to an expert-based approach, currently utilized by a leading network provider, which is presented in the following.

\subsection{Benchmark algorithm}
\label{sec:benchmark}

To benchmark the performance of our modeling framework, we employ an approach used by experts to predict carrier shutdown network performance in the field. It is crucial to note that this scheme is deterministic. In more detail, it first assumes that all capacity cells are active at the beginning of every hour, i.e. $x_c = 1$ for all $c \in {\cal C}_C$. Then, a sequential modeling approach is adopted, where the capacity cells are  shut down one by one in a given order. The average numbers, $U^{\rm UE}_{i,h}$ and $\Delta^{\rm DL}_{i,h}$, of \ac{RRC}-connected \acp{UE} to and used \ac{DL} \acp{PRB} by every cell, $i$, at each hour, $h$, are used to check the shutdown condition rule, $u_{\rm sd}$ (see Section \ref{sec:update_rule}), and decide the deterministic order of shutdown. Once this order is fixed, the capacity cells go through the \ac{UE} transfer scheme presented in Section \ref{sec:user_transfer}. After processing all capacity cells that meet the shutdown condition rule, $u_{\rm sd}$, the statistics for the hour are derived. These statistics include the predicted average numbers, $\hat{U}^{\rm UE}_{i,h^\prime}$ and $\hat{\Delta}^{\rm DL}_{i,h^\prime}$, of \ac{RRC}-connected \acp{UE} to and used \ac{DL} \acp{PRB} by every cell, $i$, at each hour, $h^\prime$, as well as the predicted duration, $\hat{t}^{\rm CS}_{i,h^\prime}$, of the carrier shutdown. These values are feed into the \ac{ML} model presented in Section \ref{sec:pc_model} to provide an estimate of cell power consumption. As for the \ac{UE} rate prediction, the mean estimator presented in Section \ref{sec:rate_model} is used. It's important to note that this method only allows obtaining one estimate for each target cell \ac{KPI} in each hour, not a distribution, as the model is not stochastic, in contrast with the proposed framework.

\subsection{Convergence analysis of ABM estimation errors}
\label{sec:error analysis}

\begin{figure}[th!]
    \centering
    \includegraphics[clip,scale=0.7]
    {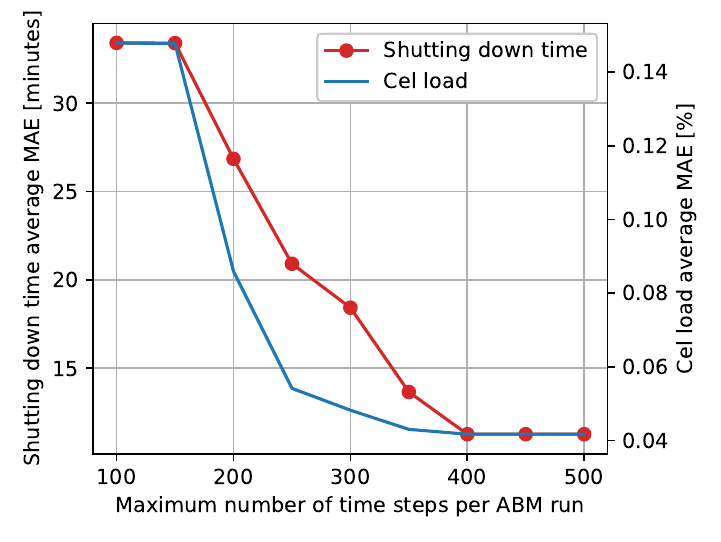}
    \caption{Average MAEs with respect to the maximum number of time steps per ABM iteration.}
    \label{fig:error_trend}
\end{figure}

To validate the capabilities of the proposed model, 
this section analyzes the convergence of estimation errors as we increase the maximum number of time steps allowed in each Monte Carlo run of the ABM. 
Fig. \ref{fig:error_trend} demonstrates that the average MAE of the estimates produced by the ABM is significantly influenced by this parameter. 
Notably, when this threshold is extended beyond 150, 
both types of errors rapidly decrease, 
achieving convergence once the maximum number of steps per Monte Carlo run reaches 400. 
By setting the maximum number of steps per Monte Carlo run to this level, 
ABM requires approximately 2 minutes to simulate network performance over a 24-hour period. 
This outcome lays the groundwork for discussing SRCON modelling performance in subsequent sections.
 
\subsection{Predicting the shutting down time of capacity cells}
\label{sec:Shutting Down Time}

In this section, we evaluate the performance of the proposed framework in estimating the carrier shutdown duration of the capacity booster cells. 

As explained in Section \ref{sec:carriershutdown}, this process is stochastic, influenced by the number of \acp{UE} associated with the shutting down capacity cell and its paired coverage cell, as well as their \ac{DL} and \ac{UL} \ac{PRB} loads. Accurately estimating the capacity cell shutdown duration is a prerequisite to assess network load and \ac{UE} performance. With respect to energy savings, it holds significance for a precise evaluation of network energy consumption, and enables an understanding of the effectiveness of the deployed energy-saving policy.

\begin{figure} [!t]
    \centering
    \includegraphics[width=8.7cm]{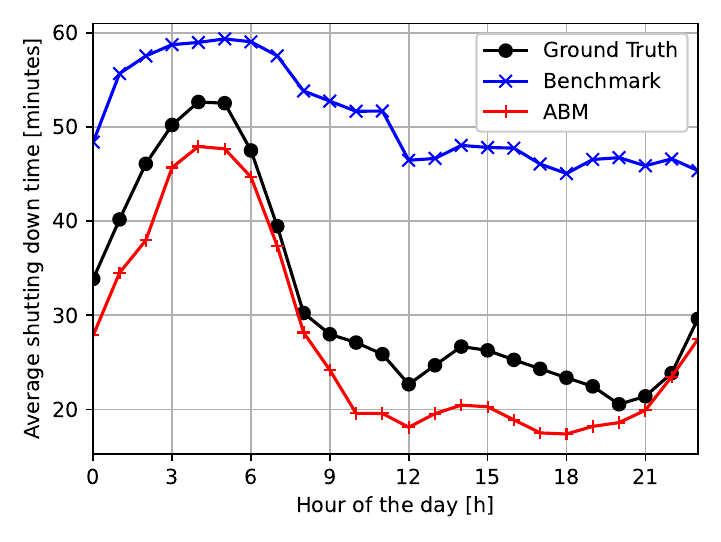}
    \caption{Daily profile of the mean shutting down time of capacity booster cells; ground truth vs the Benchmark algorithm and \ac{ABM} estimations.}
    \label{fig:Shutdown_time}
\end{figure}

Fig. \ref{fig:Shutdown_time} illustrates the daily profile of the mean shutdown time of the capacity cells in the network under consideration in the test dataset. As expected, the shutdown time increases during the night until a peak, after which the network load begins to rise, causing the mean shutdown time to decrease accordingly. In the evening, as the network load reduces, additional capacity cells can shut down.

Additionally, it is evident from Fig. \ref{fig:Shutdown_time} that the proposed framework accurately characterizes the average shutdown time during each hour of the day by adapting its estimate to the variation in network load. In contrast, the benchmark algorithm consistently overestimates the shutdown time and fails to adjust its estimate according to the network load, this remains rather flat throughout the day.

Table \ref{Table:shutdown time accuracy} describes the average ${\ac{MAE}}$, and ${\ac{MAPE}}$ achieved by the investigated algorithms and highlights the accuracy improvement of \ac{ABM} with respect to the Benchmark algorithm. Specifically, over 24 hours, 
we observe accuracy gain of 48.53\%, and 87.28\%, in terms of $\overline{\ac{MAE}}$, and $\overline{\ac{MAPE}}$, respectively.

\begin{table}[h]
\caption{{Estimation accuracy of the carrier shutdown time of the capacity booster cells.}}
\scriptsize
\centering
\begin{tabular}{l|cccc}
{Algorithms}            & $\overline{\ac{MAE}}$             & $\overline{\ac{MAPE}}$  \\ \hline\hline
\textit{\ac{ABM}}       &  11.27min/h                       & 101\% \\ \hline
\textit{Benchmark}      & 21.90min/h                        & 794\%    \\
                             \hline
\textit{Accuracy Gain}  & 48.53\%                           & 87.28\% \\                        \hline\hline
\end{tabular}
\label{Table:shutdown time accuracy}
\end{table}

\subsection{Predicting the load of network cells}
\label{sec:Cell_load}

In this section, we assess the performance of the proposed framework in estimating the cell load.

At the network level, load variation is a well-known phenomenon, decreasing at night and increasing in the morning until reaching a peak in the late evening. At the cell level, in the absence of energy-saving features, load variation primarily depends on user mobility. However, the activation of carrier shutdown introduces a new dynamic, as lightly loaded capacity booster cells attempt to handover their \acp{UE} (as explained in Sec. \ref{sec:ue_transfer_logic}) to switch off and reduce network energy consumption. Consequently, carrier shutdown also impacts the cell load distribution in the network.

\begin{figure} [!t]
    \centering
    \includegraphics[width=8.7cm]{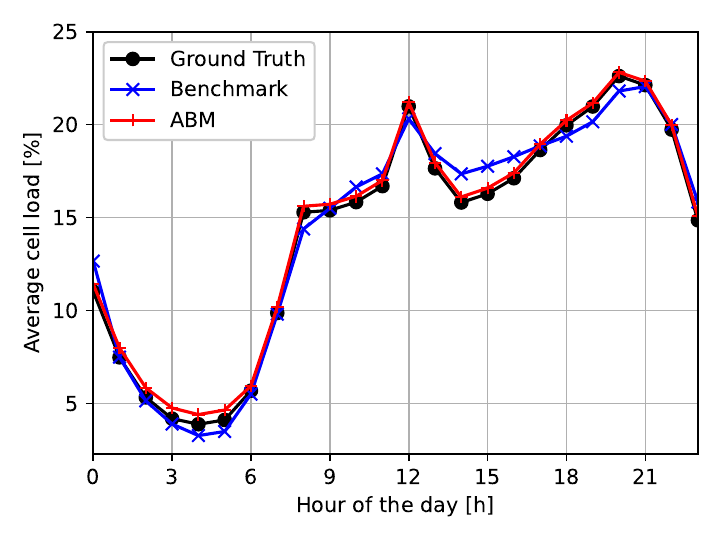}
    \caption{Daily profile of the average cell load; ground truth vs the Benchmark algorithm and \ac{ABM} estimations.}
    \label{fig:Cell_load}
\end{figure}

Fig. \ref{fig:Cell_load} illustrates the daily profile of the average cell load in the considered network. As expected, its pattern aligns with the typical network load profile described in \cite{Auer2011}. This plot demonstrates how both the estimates of the proposed framework and the benchmark follow the trend of the network load during the day. Our experiments reveal that the \ac{ABM} estimate adeptly follows drastic variations in cell load profiles, providing more precise estimations than the benchmark algorithm. Specifically, as summarized in Table \ref{Table:cell_load}, the \ac{ABM} improves the accuracy of mean cell load predictions with respect to the benchmark algorithm, achieving a $57.89\,\%$ and $50.74\,\%$ reduction in terms of average \ac{MAE} and \ac{MAPE}, respectively.

\begin{table}[h]
\caption{{Estimation accuracy of the cell load.}}
\scriptsize
\centering
\begin{tabular}{l|cccc}
{Algorithms} & $\overline{\ac{MAE}}$           & $\overline{\ac{MAPE}}$  \\ \hline\hline
\textit{\ac{ABM}}     &  4.17\%     & 49.4\% \\ \hline
\textit{Benchmark}             & 9.9\%    & 100.3\%    \\
                             \hline
\textit{Accuracy Gain}              & 57.89\%      & 50.74\% \\                        \hline\hline
\end{tabular}
\label{Table:cell_load}
\end{table}

\subsection{Predicting the energy consumption of capacity cells}
\label{sec:energy_consumption}

\begin{figure} [!t]
    \centering
    \includegraphics[width=8.7cm]{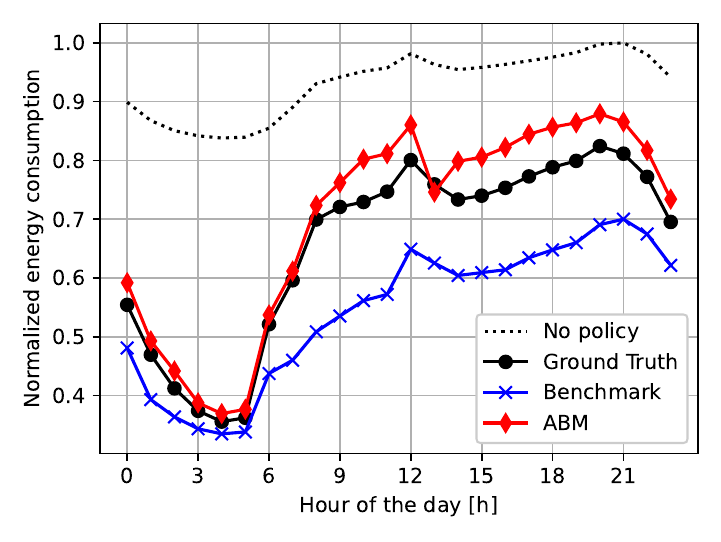}
    \caption{Daily profile of the energy consumption of capacity booster cells; ground truth vs the Benchmark algorithm and \ac{ABM} estimations.}
    \label{fig:energy_consumption}
\end{figure}

In this section, we assess the performance of the proposed \ac{ABM} in estimating the energy consumption of the capacity cells.
Fig.~\ref{fig:energy_consumption} depicts the ground-truth average energy consumed by all the capacity cells at each hour of the day, 
both with and without activating the energy-saving policy, i.e. in the test and training datasets. 
It is important to note that the values have been normalized for privacy reasons.

When carrier shutdown is active, the power consumption of the capacity cells is high during the daytime, reaching its peak at 8 pm, and then decreases during the night until 4 am when it reaches its minimum value. In contrast, the power consumption profile of the capacity cells is notably flatter when the energy-saving policy is not active, i.e. without carrier shutdown, the energy consumption does not scale well with the actual network load. From Fig.\ref{fig:energy_consumption}, it is observed that, on average, adjusting the network capacity to the variations of network load through energy-saving policies results in an energy consumption saving of $50\,\%$ at the capacity cells. 

The power consumption estimations computed through the power consumption model presented in Section \ref{sec:pc_model}, using the estimated shutdown times and loads of the cells in the network based on the proposed \ac{ABM} model and the benchmark scheme, are also reported in the figure. Notably, both algorithms successfully capture the general trend of energy consumption. However, the estimation based on the proposed \ac{ABM} model is much closer to reality, owing to its superior capability in accurately estimating the mentioned loads and shutdown times.

Table~\ref{Table:energy_consumption} details the average \ac{MAE} and \ac{MAPE} achieved by the investigated algorithms, showcasing the accuracy improvement of \ac{ABM} compared to the benchmark algorithm. Specifically, over 24 hours, we observe a significant accuracy gain of 62.08\,\% and 62.06\,\% in terms of the average \ac{MAE} and \ac{MAPE}, respectively.

\begin{table}[h]
\caption{{Estimation accuracy of the energy consumption of the capacity booster cells.}}
\scriptsize
\centering
\begin{tabular}{l|cccc}
{Algorithms}            & $\overline{\ac{MAE}}$         & $\overline{\ac{MAPE}}$  \\ \hline\hline
\textit{\ac{ABM}}       &  33.52 Wh                     &  6.27\% \\ \hline
\textit{Benchmark}      & 88.43 Wh                      & 16.54\%    \\
                             \hline
\textit{Accuracy Gain}  & 62.08\%                       & 62.06\% \\                        \hline\hline
\end{tabular}
\label{Table:energy_consumption}
\end{table}

\subsection{Predicting the UE DL Rate}
\label{sec:ue_dl_rate}

\begin{figure} [!t]
    \centering
    \includegraphics[width=8.7cm]{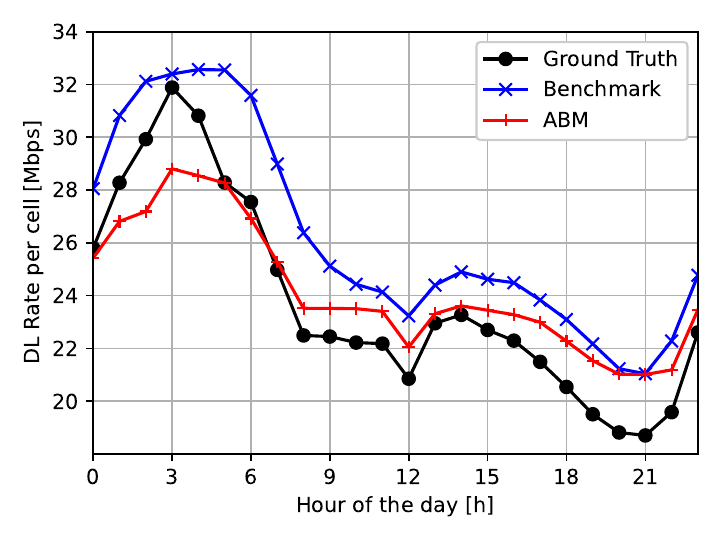}
    \caption{Daily profile of the average user rate; ground truth vs the Benchmark algorithm and \ac{ABM} estimations.}
    \label{fig:user_rate}
\end{figure}

To conclude our analysis, we discuss the capabilities of our framework in characterizing the mean \ac{UE} \ac{DL} rate, as illustrated in Fig. \ref{fig:user_rate}. As expected, this parameter exhibits an opposite trend to the cell load, detailed in Fig. \ref{fig:Cell_load}. During the night, when the cell load decreases, more resources become available in each cell for active \acp{UE}, leading to a better end-user experience. In contrast, as the cell load increases during the day, the corresponding user rate decreases due to the limited resources per active \ac{UE}.

From these results, we can observe that the proposed \ac{UE} rate estimator based on gradient boosting presented in Section \ref{sec:rate_model} tends to underestimate the true data rate during nighttime and slightly overestimate it during the daytime. In contrast, the benchmark estimator consistently overestimates the ground truth, resulting in larger errors compared to the proposed model. The proposed framework, building on our \ac{ABM}, can account for the impact of the energy-saving scheme on the cell-level \acp{KPI}, which ultimately determines the \ac{UE} rate.

As summarized in Table \ref{Table:user_rate}, our solution significantly improves the accuracy of the benchmark when estimating the user rate, achieving a nearly $50\,\%$ reduction in mean \ac{MAE} and around a $57\,\%$ reduction in mean \ac{MAPE}.

\begin{table}[h]
\caption{{Estimation accuracy of the user rate.}}
\scriptsize
\centering
\begin{tabular}{l|cccc}
{Algorithms}            & $\overline{\ac{MAE}}$             & $\overline{\ac{MAPE}}$  \\ \hline\hline
\textit{\ac{ABM}}       &  3.92 Mbps                        &  20.03\% \\ \hline
\textit{Benchmark}      & 7.72 Mbps                         & 46.44\%    \\
                             \hline
\textit{Accuracy Gain}  & 49.19\%                           & 56.86\% \\                        \hline\hline
\end{tabular}
\label{Table:user_rate}
\end{table}

{\color {black}
\subsection{Discussion on generalization to other scenarios}

It is imperative to note that the precise calibration of our model is fundamentally intertwined with the specificities of the vendor's equipment and the deployment characteristics it was originally designed for. Consequently, 
the model presented in this paper exhibits an optimized performance for metropolitan networks that deploy analogous technological frameworks than those used in training scenario, specifically 4G and 5G networks. 

The necessity for comprehensive retraining of the \ac{ML} models arises when attempting to transpose our energy-saving solutions and performance prediction models to alternative network setups or deployment scenarios. This retraining process is vital for accommodating the varying power consumption profiles and environmental conditions unique to different equipment or scenarios.

Furthermore, deviations in the carrier shutdown and \ac{HO} algorithms, from the framework specified, necessitate a thoughtful redesign of the corresponding expert-based models within our \ac{ABM}. Such adaptations underscore our commitment to providing precise, localized predictions. 

This deliberation accentuates the importance of a thorough evaluation of network infrastructure, vendor-specific technologies, and the nuanced characteristics of the network environment prior to the deployment of our proposed model in real-world settings.
}
\section{Conclusions}
\label{sec:conclusions}

{\color {black} 
In conclusion, 
this paper not only introduced the \ac{SRCON} framework, 
but also rigorously evaluated its capabilities in the context of the research questions and scientific objectives laid out at the outset. 
Through this comprehensive data-driven approach, 
aimed at modeling the energy efficiency of cellular networks with a focus on carrier shutdown policies, 
we have made significant strides towards addressing the pressing challenges identified in our research.

\textbf{Feasibility of Data-Driven Modeling:}
Our findings affirmatively answer the first research question, 
demonstrating that a data-driven model, 
relying exclusively on readily accessible network data and excluding costly \ac{DT}, 
can indeed deliver accurate predictions of network energy consumption and \ac{UE} throughput. 
The \ac{SRCON} framework, 
by leveraging advanced \ac{ML}- and expert-based models, 
has accurately estimated key network performance parameters in response to diverse carrier shutdown configurations, 
thereby showcasing remarkable accuracy.

\textbf{Blending Machine Learning with Domain Expertise:}
In addressing our second research question, 
the \ac{SRCON} framework exemplifies the seamless integration of \ac{ML} algorithms with domain expertise. 
Our innovative, expert-driven \ac{ABM} incorporates expert knowledge by implementing product algorithms, 
specifically focusing on carrier shutdown and \ac{HO} mechanisms. 
It also effectively interacts with our novel \ac{ML}-based models for energy consumption and \ac{UE} throughput to ensure precise predictions. 
This integration has resulted in significant accuracy improvements, 
underscoring the efficacy of our approach in advancing energy efficiency modeling within large-scale networks.

\textbf{Advancements and Comparative Evaluation:}
Regarding our third question, 
the \ac{SRCON} framework has introduced notable advancements in network energy efficiency prediction.
When compared to existing methodologies, 
notably a state-of-the-art approach used by a network operator, \ac{SRCON} achieved a 62.08\,\% reduction in \ac{MAE} and a 62.06\,\% reduction in \ac{MAPE} for energy consumption modeling. 
For \ac{UE} throughput estimation, 
significant improvements were observed with approximately a 50\,\% reduction in \ac{MAE} and a 57\,\% reduction in \ac{MAPE}, 
underscoring its superior predictive performance.

In summary, 
the \ac{SRCON} framework signifies a paradigm shift towards practical, data-driven network modeling and optimization. 
It leverages existing network data to accurately predict resource utilization, energy consumption, and \ac{UE} throughput, 
offering a viable solution for modeling and optimizing wide-area networks and fulfilling both the research questions posed and the scientific objectives set forth at the beginning of this study.}

\section{Acknowledgement}

This research is supported in part by the NSF project CNS 1910153, the Generalitat Valenciana through the CIDEGENT PlaGenT, Grant CIDEXG/2022/17, Project iTENTE, and by the action CNS2023-144333, financed by MCIN/AEI/10.13039/501100011033 and the European Union “NextGenerationEU”/PRTR.”




\balance
\bibliographystyle{IEEEtran}
\bibliography{bibl.bib}

\begin{acronym}[AAAAAAAAA]  
\acro{2D}{two-dimensional}
 \acro{3D}{three-dimensional}
 \acro{3G}{third generation}
 \acro{3GPP}{third generation partnership project}
 \acro{4G}{fourth generation}
 \acro{5G}{fifth generation}
 \acro{5GC}{5G core network}
 \acro{AAA}{authentication, authorisation and accounting}
 \acro{ABM}{agent-based model}
 \acro{ABS}{almost blank subframe}
 \acro{AC}{alternating current}
 \acro{ACIR}{adjacent channel interference rejection ratio}
 \acro{ACK}{acknowledgment}
 \acro{ACL}{allowed CSG list}
 \acro{ACLR}{adjacent channel leakage ratio}
 \acro{ACPR}{adjacent channel power ratio}
 \acro{ACS}{adjacent channel selectivity}
 \acro{ADC}{analog-to-digital converter}
 \acro{ADSL}{asymmetric digital subscriber line}
 \acro{AEE}{area energy efficiency}
 \acro{AF}{amplify-and-forward}
 \acro{AGCH}{access grant channel}
 \acro{AGG}{aggressor cell}
 \acro{AH}{authentication header}
 \acro{AI}{artificial intelligence}
 \acro{AKA}{authentication and key agreement}
 \acro{AMC}{adaptive modulation and coding}
 \acro{ANN}{artificial neural network}
 \acro{ANR}{automatic neighbour relation}
 \acro{AoA}{angle of arrival}
 \acro{AoD}{angle of departure}
 \acro{APC}{area power consumption}
 \acro{API}{application programming interface}
 \acro{APP}{a posteriori probability}
 \acro{AR}{augmented reality}
 \acro{ARIMA}{autoregressive integrated moving average}
 \acro{ARQ}{automatic repeat request}
 \acro{AS}{access stratum}
 \acro{ASE}{area spectral efficiency}
 \acro{ASM}{advanced sleep mode}
 \acro{ASN}{access service network}
 \acro{ATM}{asynchronous transfer mode}
 \acro{ATSC}{Advanced Television Systems Committee}
 \acro{AUC}{authentication centre}
 \acro{AWGN}{additive white gaussian noise}
 \acro{BB}{baseband}
 \acro{BBU}{baseband unit}
 \acro{BCCH}{broadcast control channel}
 \acro{BCH}{broadcast channel}
 \acro{BCJR}{Bahl-Cocke-Jelinek-Raviv} 
 \acro{BE}{best effort}
 \acro{BER}{bit error rate}
 \acro{BLER}{block error rate}
 \acro{BPSK}{binary phase-shift keying}
 \acro{BR}{bit rate}
 \acro{BS}{base station}
 \acro{BSC}{base station controller}
 \acro{BSIC}{base station identity code}
 \acro{BSP}{binary space partitioning}
 \acro{BSS}{blind source separation}
 \acro{BTS}{base transceiver station}
 \acro{BWP}{bandwidth part}
 \acro{CA}{carrier aggregation}
 \acro{CAC}{call admission control}
 \acro{CaCo}{carrier component}
 \acro{CAPEX}{capital expenditure}
 \acro{capex}{capital expenses}
 \acro{CAS}{cluster angular spread}
 \acro{CATV}{community antenna television}
 \acro{CAZAC}{constant amplitude zero auto-correlation}
 \acro{CC}{component carrier}
 \acro{CCCH}{common control channel}
 \acro{CCDF}{complementary cumulative distribution function}
 \acro{CCE}{control channel element}
 \acro{CCO}{coverage and capacity optimisation}
 \acro{CCPCH}{common control physical channel}
 \acro{CCRS}{coordinated and cooperative relay system}
 \acro{CCTrCH}{coded composite transport channel}
 \acro{CDF}{cumulative distribution function}
 \acro{CDMA}{code division multiple access}
\acro{CDR}{call detail record}
 \acro{CDS}{cluster delay spread}
 \acro{CESM}{capacity effective SINR mapping}
 \acro{CO$_{2e}$}{carbon dioxide equivalent}
 \acro{CFI}{control format indicator}
 \acro{CFL}{Courant-Friedrichs-Lewy}
 \acro{CGI}{cell global identity}
 \acro{CID}{connection identifier}
 \acro{CIF}{carrier indicator field}
 \acro{CIO}{cell individual offset}
 \acro{CIR}{channel impulse response}
 \acro{CNN}{convolutional neural network}
 \acro{CMF}{cumulative mass function}
 \acro{C-MIMO}{cooperative MIMO}
 \acro{CN}{core network}
 \acro{COC}{cell outage compensation}
 \acro{COD}{cell outage detection}
 \acro{CoMP}{coordinated multi-point}
 \acro{ConvLSTM}{convolutional LSTM}
 \acro{CP}{cycle prefix}
 \acro{CPC}{cognitive pilot channel}
 \acro{CPCH}{common packet channel}
 \acro{CPE}{customer premises equipment}
 \acro{CPICH}{common pilot channel}
 \acro{CPRI}{common public radio interface}
 \acro{CPU}{central processing unit}
 \acro{CQI}{channel quality indicator}
 \acro{CR}{cognitive radio}
 \acro{CRAN}{centralized radio access network} 
 \acro{C-RAN}{cloud radio access network} 
 \acro{CRC}{cyclic redundancy check}
 \acro{CRE}{cell range expansion}
 \acro{C-RNTI}{cell radio network temporary identifier}
 \acro{CRP}{cell re-selection parameter}
 \acro{CRS}{cell-specific reference symbol}
 \acro{CRT}{cell re-selection threshold}
 \acro{CSCC}{common spectrum coordination channel}
 \acro{CSG ID}{closed subscriber group ID}
 \acro{CSG}{closed subscriber group}
 \acro{CSI}{channel state information}
 \acro{CSIR}{receiver-side channel state information}
 \acro{CSI-RS}{channel state information-reference signals}
 \acro{CSO}{cell selection offset}
 \acro{CTCH}{common traffic channel}
 \acro{CTS}{clear-to-send} 
 \acro{CU}{central unit}
 \acro{CV}{cross-validation}
 \acro{CWiND}{Centre for Wireless Network Design}
 \acro{D2D}{device to device}
 \acro{DAB}{digital audio broadcasting}
 \acro{DAC}{digital-to-analog converter}
 \acro{DAS}{distributed antenna system}
 \acro{dB}{decibel}
 \acro{dBi}{isotropic-decibel}
 \acro{DC}{direct current}
 \acro{DCCH}{dedicated control channel}
 \acro{DCF}{decode-and-forward}
 \acro{DCH}{dedicated channel}
 \acro{DC-HSPA}{dual-carrier high speed packet access}
 \acro{DCI}{downlink control information}
 \acro{DCM}{directional channel model}
 \acro{DCP}{dirty-paper coding}
 \acro{DCS}{digital communication system}
 \acro{DECT}{digital enhanced cordless telecommunication}
 \acro{DeNB}{donor eNodeB}
 \acro{DFP}{dynamic frequency planning}
 \acro{DFS}{dynamic frequency selection}
 \acro{DFT}{discrete Fourier transform}
 \acro{DFTS}{discrete Fourier transform spread}
 \acro{DHCP}{dynamic host control protocol}
 \acro{DL}{downlink}
 \acro{DMC}{dense multi-path components}
 \acro{DMF}{demodulate-and-forward}
 \acro{DMT}{diversity and multiplexing tradeoff}
  \acro{DNN}{deep neural network} 
 \acro{DoA}{direction-of-arrival}
 \acro{DoD}{direction-of-departure}
 \acro{DoS}{denial of service}
 \acro{DPCCH}{dedicated physical control channel}
 \acro{DPDCH}{dedicated physical data channel}
 \acro{D-QDCR}{distributed QoS-based dynamic channel reservation}
 \acro{DQL}{deep Q-learning}
\acro{DRAN}{distributed radio access network}
 \acro{DRS}{discovery reference signal}
 \acro{DRL}{deep reinforcement learning}
 \acro{DRX}{discontinuous reception}
 \acro{DS}{down stream}
 \acro{DSA}{dynamic spectrum access}
 \acro{DSCH}{downlink shared channel}
 \acro{DSL}{digital subscriber line}
 \acro{DSLAM}{digital subscriber line access multiplexer}
 \acro{DSP}{digital signal processor}
 \acro{DT}{drive test}
 \acro{DTCH}{dedicated traffic channel}
 \acro{DTX}{discontinuous transmission}
   \acro{DU}{distributed unit}
 \acro{DVB}{digital video broadcasting}
 \acro{DXF}{drawing interchange format}
 \acro{E2E}{end-to-end}
 \acro{EAGCH}{enhanced uplink absolute grant channel}
 \acro{EA}{evolutionary algorithm}
 \acro{EAP}{extensible authentication protocol}
 \acro{EC}{evolutionary computing}
 \acro{ECGI}{evolved cell global identifier}
 \acro{ECR}{energy consumption ratio}
 \acro{ECRM}{effective code rate map}
 \acro{EDCH}{enhanced dedicated channel}
 \acro{EE}{energy efficiency}
 \acro{EESM}{exponential effective SINR mapping}
 \acro{EF}{estimate-and-forward}
 \acro{EGC}{equal gain combining}
 \acro{EHICH}{EDCH HARQ indicator channel}
 \acro{eICIC}{enhanced intercell interference coordination}
 \acro{EIR}{equipment identity register}
 \acro{EIRP}{effective isotropic radiated power}
 \acro{ELF}{evolutionary learning of fuzzy rules}
 \acro{eMBB}{enhanced mobile broadband}
  \acro{EMR}{Electromagnetic-Radiation}
 \acro{EMS}{enhanced messaging service}
 \acro{eNB}{evolved NodeB}
 \acro{eNodeB}{evolved NodeB}
 \acro{EoA}{elevation of arrival}
 \acro{EoD}{elevation of departure}
 \acro{EPB}{equal path-loss boundary}
 \acro{EPC}{evolved packet core}
 \acro{EPDCCH}{enhanced physical downlink control channel}
 \acro{EPLMN}{equivalent PLMN}
 \acro{EPS}{evolved packet system}
 \acro{ERAB}{eUTRAN radio access bearer}
 \acro{ERGC}{enhanced uplink relative grant channel}
 \acro{ERTPS}{extended real time polling service}
 \acro{ESB}{equal downlink receive signal strength boundary}
 \acro{ESF}{even subframe}
 \acro{ESP}{encapsulating security payload}
 \acro{ETSI}{European telecommunications standards institute}
 \acro{E-UTRA}{evolved UTRA}
 \acro{EU}{European Union}
 \acro{EUTRAN}{evolved UTRAN}
 \acro{EVDO}{evolution-data optimised}
 \acro{FACCH}{fast associated control channel}
 \acro{FACH}{forward access channel}
 \acro{FAP}{femtocell access point}
 \acro{FARL}{fuzzy assisted reinforcement learning}
 \acro{FCC}{Federal Communications Commission}
 \acro{FCCH}{frequency-correlation channel}
 \acro{FCFS}{first-come first-served}
 \acro{FCH}{frame control header}
 \acro{FCI}{failure cell ID}
 \acro{FD}{frequency-domain}
 \acro{FDD}{frequency division duplexing}
 \acro{FDM}{frequency division multiplexing}
 \acro{FDMA}{frequency division multiple access}
 \acro{FDTD}{finite-difference time-domain}
 \acro{FE}{front-end}
 \acro{FeMBMS}{further evolved multimedia broadcast multicast service}
 \acro{FER}{frame error rate}
 \acro{FFR}{fractional frequency reuse}
 \acro{FFRS}{fractional frequency reuse scheme}
 \acro{FFT}{fast Fourier transform}
 \acro{FFU}{flexible frequency usage}
 \acro{FGW}{femtocell gateway}
 \acro{FIFO}{first-in first-out}
 \acro{FIS}{fuzzy inference system}
 \acro{FMC}{fixed mobile convergence}
 \acro{FPC}{fractional power control}
 \acro{FPGA}{field-programmable gate array}
 \acro{FRS}{frequency reuse scheme}
 \acro{FTP}{file transfer protocol}
 \acro{FTTx}{fiber to the x}
 \acro{FUSC}{full usage of subchannels}
 \acro{GA}{genetic algorithm}
 \acro{GAN} {generative adversarial network}
 \acro{GANC}{generic access network controller}
 \acro{GBR}{guaranteed bitrate}
 \acro{GCI}{global cell identity}
 \acro{GERAN}{GSM edge radio access network}
 \acro{GGSN}{gateway GPRS support node}
 \acro{GHG}{greenhouse gas}
 \acro{GMSC}{gateway mobile switching centre}
 \acro{gNB}{next generation NodeB}
 \acro{GNN}{graph neural network}
 \acro{GCN}{graph convolutional network}
 \acro{GNSS}{global navigation satellite system}
 \acro{GP}{genetic programming}
 \acro{GPON}{Gigabit passive optical network}
 \acro{GPP}{general purpose processor}
 \acro{GPRS}{general packet radio service}
 \acro{GPS}{global positioning system}
 \acro{GPU}{graphics processing unit}
 \acro{GRU}{gated recurrent unit}
 \acro{GSCM}{geometry-based stochastic channel models}
 \acro{GSMA}{global system for mobile communications association}
 \acro{GSM}{global system for mobile communication}
 \acro{GTD}{geometry theory of diffraction}
 \acro{GTP}{GPRS tunnel protocol}
 \acro{GTP-U}{GPRS tunnel protocol - user plane}
 \acro{HA}{hybrid access}
 \acro{HARQ}{hybrid automatic repeat request}
 \acro{HBS}{home base station}
 \acro{HCN}{heterogeneous cellular network}
 \acro{HCS}{hierarchical cell structure}
  \acro{HD}{high definition}
 \acro{HDFP}{horizontal dynamic frequency planning}
 \acro{HeNB}{home eNodeB}
 \acro{HeNodeB}{home eNodeB}
 \acro{HetNet}{heterogeneous network}
 \acro{HiFi}{high fidelity}
 \acro{HII}{high interference indicator}
 \acro{HLR}{home location register}
 \acro{HNB}{home NodeB}
 \acro{HNBAP}{home NodeB application protocol}
 \acro{HNBGW}{home NodeB gateway}
 \acro{HNodeB}{home NodeB}
 \acro{HO}{handover}
 \acro{HOF}{handover failure}
 \acro{HOM}{handover hysteresis margin}
 \acro{HPBW}{half power beam width}
 \acro{HPLMN}{home PLMN}
 \acro{HPPP}{homogeneous Poison point process}
 \acro{HRD}{horizontal reflection diffraction}
 \acro{HSB}{hot spot boundary}
 \acro{HSDPA}{high speed downlink packet access}
 \acro{HSDSCH}{high-speed DSCH}
 \acro{HSPA}{high speed packet access}
 \acro{HSS}{home subscriber server}
 \acro{HSUPA}{high speed uplink packet access}
 \acro{HUA}{home user agent}
 \acro{HUE}{home user equipment}
 \acro{HVAC}{heating, ventilating, and air conditioning}
 \acro{IC}{interference cancellation}
 \acro{ICI}{inter-carrier interference}
 \acro{ICIC}{intercell interference coordination}
 \acro{ICNIRP}{International Commission on Non-Ionising Radiation Protection}
 \acro{ICS}{IMS centralised service}
 \acro{ICT}{information and communication technology}
 \acro{ID}{identifier}
 \acro{IDFT}{inverse discrete Fourier transform}
 \acro{IE}{information element}
 \acro{IEEE}{institute of electrical and electronics engineers}
 \acro{IETF}{internet engineering task force}
 \acro{IFA}{inverted-F-antennas}
 \acro{IFFT}{inverse fast Fourier transform}
 \acro{i.i.d.}{independent and identical distributed}
 \acro{IIR}{infinite impulse response}
 \acro{IKE}{Internet key exchange}
 \acro{IKEv2}{Internet key exchange version 2}
 \acro{ILP}{integer linear programming}
 \acro{IMEI}{international mobile equipment identity}
 \acro{IMS}{IP multimedia subsystem}
 \acro{IMSI}{international mobile subscriber identity}
 \acro{IMT}{international mobile telecommunications}
 \acro{INH}{indoor hotspot}
 \acro{IOI}{interference overload indicator}
 \acro{IoT}{Internet of things}
 \acro{IP}{Internet protocol}
 \acro{IPSEC}{Internet protocol security}
 \acro{IR}{incremental redundancy}
 \acro{IRC}{interference rejection combining}
 \acro{ISD}{inter site distance}
 \acro{ISI}{inter symbol interference}
 \acro{ITU}{international telecommunication union}
 \acro{Iub}{UMTS interface between RNC and NodeB}
 \acro{IWF}{IMS interworking function}
 \acro{JFI}{Jain's fairness index}
 \acro{KPI}{key performance indicator}
 \acro{KNN}{$k$-nearest neighbours}
 \acro{L1}{layer one}
 \acro{L2}{layer two}
 \acro{L3}{layer three}
 \acro{LA}{location area}
 \acro{LAA}{licensed Assisted Access}
 \acro{LAC}{location area code}
 \acro{LAI}{location area identity}
 \acro{LAU}{location area update}
 \acro{LDA}{linear discriminant analysis} 
 \acro{LIDAR}{laser imaging detection and ranging}
 \acro{LLR}{log-likelihood ratio}
 \acro{LLS}{link-level simulation}
 \acro{LMDS}{local multipoint distribution service}
 \acro{LMMSE}{linear minimum mean-square-error}
 \acro{LoS}{line-of-sight}
 \acro{LPC}{logical PDCCH candidate}
 \acro{LPN}{low power node}
 \acro{LR}{likelihood ratio}
 \acro{LSAS}{large-scale antenna system}
 \acro{LSP}{large-scale parameter}
 \acro{LSTM}{long short term memory cell}
 \acro{LTE/SAE}{long term evolution/system architecture evolution}
 \acro{LTE}{long term evolution}
 \acro{LTE-A}{long term evolution advanced}
 \acro{LUT}{look up table}
 \acro{MAC}{medium access control}
 \acro{MaCe}{macro cell}
 \acro{MAE}{mean absolute error}
 \acro{MAP}{media access protocol}
 \acro{MAPE}{mean absolute percentage error}
 \acro{MAXI}{maximum insertion}
 \acro{MAXR}{maximum removal}
 \acro{MBMS}{multicast broadcast multimedia service} 
 \acro{MBS}{macrocell base station}
 \acro{MBSFN}{multicast-broadcast single-frequency network}
 \acro{MC}{modulation and coding}
 \acro{MCB}{main circuit board}
 \acro{MCM}{multi-carrier modulation}
 \acro{MCP}{multi-cell processing}
 \acro{MCPA}{multi-carrier power amplifier}
 \acro{MCS}{modulation and coding scheme}
 \acro{MCSR}{multi-carrier soft reuse}
 \acro{MDAF}{management data analytics function}
 \acro{MDP}{markov decision process }
 \acro{MDT}{minimisation of drive tests}
 \acro{MEA}{multi-element antenna}
 \acro{MeNodeB}{Master eNodeB}
 \acro{MGW}{media gateway}
 \acro{MIB}{master information block}
 \acro{MIC}{mean instantaneous capacity}
 \acro{MIESM}{mutual information effective SINR mapping}
 \acro{MIMO}{multiple-input multiple-output}
 \acro{MINI}{minimum insertion}
 \acro{MINR}{minimum removal}
 \acro{MIP}{mixed integer program}
 \acro{MISO}{multiple-input single-output}
 \acro{ML}{machine learning}
 \acro{MLB}{mobility load balancing}
 \acro{MLB}{mobility load balancing}
 \acro{MM}{mobility management}
 \acro{MME}{mobility management entity}
 \acro{mMIMO}{massive multiple-input multiple-output}
 \acro{MMSE}{minimum mean square error}
 \acro{mMTC}{massive machine type communication}
 \acro{MNC}{mobile network code}
 \acro{MNO}{mobile network operator}
 \acro{MOS}{mean opinion score}
 \acro{MPC}{multi-path component}
 \acro{MR}{measurement report}
 \acro{MRC}{maximal ratio combining}
 \acro{MR-FDPF}{multi-resolution frequency-domain parflow}
 \acro{MRO}{mobility robustness optimisation}
 \acro{MRT}{maximum ratio transmission}
 \acro{MS}{mobile station}
 \acro{MSC}{mobile switching centre}
 \acro{MSE}{mean square error}
 \acro{MSISDN}{mobile subscriber integrated services digital network number}
 \acro{MUE}{macrocell user equipment}
 \acro{MU-MIMO}{multi-user MIMO}
 \acro{MVNO}{mobile virtual network operators}
 \acro{NACK}{negative acknowledgment}
 \acro{NAS}{non access stratum}
 \acro{NAV}{network allocation vector}
 \acro{NB}{Naive Bayes}   
 \acro{NCL}{neighbour cell list}
 \acro{NEE}{network energy efficiency}
  \acro{NF}{network function}
 \acro{NFV}{network functions virtualization}
 \acro{NG}{next generation}
 \acro{NGMN}{next generation mobile networks}
 \acro{NG-RAN}{next generation radio access network} 
 \acro{NIR}{non ionisation radiation}
 \acro{NLoS}{non-line-of-sight}
 \acro{NN}{nearest neighbour} 
 \acro{NR}{new radio}
 \acro{NRMSE}{normalised root mean square error}
 \acro{NRTPS}{non-real-time polling service}
 \acro{NSS}{network switching subsystem}
 \acro{NTP}{network time protocol}
 \acro{NWG}{network working group}
 \acro{NWDAF}{network data analytics function} 
 \acro{OA}{open access}
 \acro{OAM}{operation, administration and maintenance}
 \acro{OC}{optimum combining}
 \acro{OCXO}{oven controlled oscillator}
 \acro{ODA}{omdi-directional antenna} 
 \acro{ODU}{optical distribution unit}
 \acro{OFDM}{orthogonal frequency division multiplexing}
 \acro{OFDMA}{orthogonal frequency division multiple access}
 \acro{OFS}{orthogonally-filled subframe}
 \acro{OLT}{optical line termination}
 \acro{ONT}{optical network terminal}
 \acro{OPEX}{operational expenditure}
 \acro{OSF}{odd subframe}
 \acro{OSI}{open systems interconnection}
 \acro{OSS}{operation support subsystem}
 \acro{OTT}{over the top}
 \acro{P2MP}{point to multi-point}
 \acro{P2P}{point to point}
 \acro{PAPR}{peak-to-average power ratio}
 \acro{PA}{power amplifier}
 \acro{PBCH}{physical broadcast channel}
 \acro{PC}{power control}
 \acro{PCB}{printed circuit board}
 \acro{PCC}{primary carrier component}
 \acro{PCCH}{paging control channel}
 \acro{PCCPCH}{primary common control physical channel}
 \acro{PCell}{primary cell}
 \acro{PCFICH}{physical control format indicator channel}
 \acro{PCH}{paging channel}
 \acro{PCI}{physical layer cell identity}
 \acro{PCPICH}{primary common pilot channel}
 \acro{PCPPH}{physical common packet channel}
 \acro{PDCCH}{physical downlink control channel}
 \acro{PDCP}{packet data convergence protocol}
 \acro{PDF}{probability density function}
 \acro{PDSCH}{physical downlink shared channel}
 \acro{PDU}{packet data unit}
 \acro{PeNB}{pico eNodeB}
 \acro{PeNodeB}{pico eNodeB}
 \acro{PF}{proportional fair}
 \acro{PGW}{packet data network gateway}
 \acro{PGFL}{probability generating functional}
 \acro{PhD}{doctor of philosophy}
 \acro{PHICH}{physical HARQ indicator channel}
 \acro{PHY}{physical layer}
 \acro{PIC}{parallel interference cancellation}
 \acro{PKI}{public key infrastructure}
 \acro{PL}{path loss}
 \acro{PMI}{precoding-matrix indicator}
 \acro{PLMN ID}{public land mobile network identity}
 \acro{PLMN}{public land mobile network}
 \acro{PML}{perfectly matched layer}
 \acro{PMF}{probability mass function}
 \acro{PMP}{point to multi-point}
 \acro{PN}{pseudorandom noise}
 \acro{POI}{point of interest}
 \acro{PON}{passive optical network}
 \acro{POP}{point of presence}
 \acro{PP}{point process}
 \acro{PPP}{Poisson point process}
 \acro{PPT}{PCI planning tools}
 \acro{PRACH}{physical random access channel}
 \acro{PRB}{physical resource block}
 \acro{PSC}{primary scrambling code}
 \acro{PSD}{power spectral density}
 \acro{PSS}{primary synchronisation channel}
 \acro{PSTN}{public switched telephone network}
 \acro{PTP}{point to point}
 \acro{PUCCH}{Physical Uplink Control Channel}
 \acro{PUE}{picocell user equipment}
 \acro{PUSC}{partial usage of subchannels}
 \acro{PUSCH}{physical uplink shared channel}
 \acro{QAM}{quadrature amplitude modulation}
 \acro{QCI}{QoS class identifier}
 \acro{QoE}{quality of experience}
 \acro{QoS}{quality of service}
 \acro{QPSK}{quadrature phase-shift keying}
 \acro{RAB}{radio access bearer}
 \acro{RACH}{random access channel}
 \acro{RADIUS}{remote authentication dial-in user services}
 \acro{RAM}{Random Access Memory}
 \acro{RAN}{radio access network}
 \acro{RANAP}{radio access network application part}
 \acro{RAT}{radio access technology}
 \acro{RAU}{remote antenna unit}
 \acro{RAXN}{relay-aided x network}
 \acro{RB}{resource block}
 \acro{RCI}{re-establish cell id}
 \acro{RE}{resource efficiency}
 \acro{REB}{range expansion bias}
 \acro{REG}{resource element group}
 \acro{RF}{radio frequency}
  \acro{RFID}{radio frequency identification}
 \acro{RFP}{radio frequency planning}
 \acro{RI}{rank indicator}
 \acro{RL}{reinforcement learning}
 \acro{RLC}{radio link control}
 \acro{RLF}{radio link failure}
 \acro{RLM}{radio link monitoring}
 \acro{RMA}{rural macrocell}
 \acro{RMS}{root mean square}
 \acro{RMSE}{root mean square error}
 \acro{RN}{relay node}
 \acro{RNC}{radio network controller}
 \acro{RNL}{radio network layer}
 \acro{RNN}{recurrent neural network}
 \acro{RNP}{radio network planning}
 \acro{RNS}{radio network subsystem}
 \acro{RNTI}{radio network temporary identifier}
 \acro{RNTP}{relative narrowband transmit power}
 \acro{RPLMN}{registered PLMN}
 \acro{RPSF}{reduced-power subframes}
 \acro{RR}{round robin}
 \acro{RRC}{radio resource control}
 \acro{RRH}{remote radio head}
 \acro{RRM}{radio resource management}
 \acro{RS}{reference signal}
 \acro{RSC}{recursive systematic convolutional}
 \acro{RS-CS}{resource-specific cell-selection}
 \acro{RSQ}{reference signal quality}
 \acro{RSRP}{reference signal received power}
 \acro{RSRQ}{reference signal received quality}
 \acro{RSS}{reference signal strength}
 \acro{RSSI}{receive signal strength indicator}
 \acro{RTP}{real time transport}
 \acro{RTPS}{real-time polling service}
 \acro{RTS}{request-to-send}
 \acro{RTT}{round trip time}
  \acro{RU}{remote unit}
  \acro{RV}{random variable}
 \acro{RX}{receive}
 \acro{S1-AP}{s1 application protocol}
 \acro{S1-MME}{s1 for the control plane}
 \acro{S1-U}{s1 for the user plane}
 \acro{SA}{simulated annealing}
 \acro{SACCH}{slow associated control channel}
 \acro{SAE}{system architecture evolution}
 \acro{SAEGW}{system architecture evolution gateway}
 \acro{SAIC}{single antenna interference cancellation}
 \acro{SAP}{service access point}
 \acro{SAR}{specific absorption rate}
 \acro{SARIMA}{seasonal autoregressive integrated moving average}
 \acro{SAS}{spectrum allocation server}
 \acro{SBS}{super base station}
 \acro{SCC}{standards coordinating committee}
 \acro{SCCPCH}{secondary common control physical channel}
 \acro{SCell}{secondary cell}
 \acro{SCFDMA}{single carrier FDMA}
 \acro{SCH}{synchronisation channel}
 \acro{SCM}{spatial channel model}
 \acro{SCN}{small cell network}
 \acro{SCOFDM}{single carrier orthogonal frequency division multiplexing}
 \acro{SCP}{single cell processing}
 \acro{SCTP}{stream control transmission protocol}
 \acro{SDCCH}{standalone dedicated control channel}
 \acro{SDMA}{space-division multiple-access}
  \acro{SDO}{standard development organization}
 \acro{SDR}{software defined radio}
 \acro{SDU}{service data unit}
 \acro{SE}{spectral efficiency}
 \acro{SeNodeB}{secondary eNodeB}
 \acro{SFBC}{space frequency block coding}
 \acro{SFID}{service flow ID}
 \acro{SG}{signalling gateway}
 \acro{SGSN}{serving GPRS support node}
 \acro{SGW}{serving gateway}
 \acro{SHAP}{shapley additive explanations}
 \acro{SI}{system information}
 \acro{SIB}{system information block}
 \acro{SIB1}{systeminformationblocktype1}
 \acro{SIB4}{systeminformationblocktype4}
 \acro{SIC}{successive interference cancellation}
 \acro{SIGTRAN}{signalling transport}
 \acro{SIM}{subscriber identity module}
 \acro{SIMO}{single input multiple output}
 \acro{SINR}{signal to interference plus noise ratio}
 \acro{SIP}{session initiated protocol}
 \acro{SIR}{signal to interference ratio}
 \acro{SISO}{single input single output}
 \acro{SLAC}{stochastic local area channel}
 \acro{SLL}{secondary lobe level}
 \acro{SLNR}{signal to leakage interference and noise ratio}
 \acro{SLS}{system-level simulation}
  \acro{SMB}{small and medium-sized businesses}
 \acro{SmCe}{small cell}
 \acro{SMS}{short message service}
 \acro{SN}{serial number}
 \acro{SNMP}{simple network management protocol}
 \acro{SNR}{signal to noise ratio}
 \acro{SOCP}{second-order cone programming}
 \acro{SOHO}{small office/home office}
 \acro{SON}{self-organising network}
 \acro{son}{self-organising networks}
 \acro{SOT}{saving of transmissions}
 \acro{SPS}{spectrum policy server}
 \acro{SRCON}{simulated reality of communication networks} 
 \acro{SRS}{sounding reference signals}
 \acro{SS}{synchronization signal}
 \acro{SSB}{synchronization Signal/PBCH block}
 \acro{SIB1}{system information block 1} 
 \acro{SSL}{secure socket layer}
 \acro{SSMA}{spread spectrum multiple access}
 \acro{SSS}{secondary synchronisation channel}
 \acro{STA}{steepest ascent}
 \acro{STBC}{space-time block coding}
 \acro{SUI}{stanford university interim}
 \acro{SVR}{support vector regression}
 \acro{TA}{timing advance}
 \acro{TAC}{tracking area code}
 \acro{TAI}{tracking area identity}
 \acro{TAS}{transmit antenna selection}
 \acro{TAU}{tracking area update}
 \acro{TCH}{traffic channel}
 \acro{TCO}{total cost of ownership}
 \acro{TCP}{transmission control protocol}
 \acro{TCXO}{temperature controlled oscillator}
 \acro{TD}{temporal difference}
 \acro{TDD}{time division duplexing}
 \acro{TDM}{time division multiplexing}
 \acro{TDMA}{time division multiple access}
  \acro{TDoA}{time difference of arrival}
 \acro{TEID}{tunnel endpoint identifier}
 \acro{TLS}{transport layer security}
 \acro{TNL}{transport network layer}
  \acro{ToA}{time of arrival}
 \acro{TP}{throughput}
 \acro{TPC}{transmit power control}
 \acro{TPM}{trusted platform module}
 \acro{TR}{transition region}
 \acro{TRX}{transceiver}
 \acro{TS}{tabu search}
 \acro{TSG}{technical specification group}
 \acro{TTG}{transmit/receive transition gap}
 \acro{TTI}{transmission time interval}
 \acro{TTT}{time-to-trigger}
 \acro{TU}{typical urban}
 \acro{TV}{television}
 \acro{TWXN}{two-way exchange network}
 \acro{TX}{transmit}
 \acro{UARFCN}{UTRA absolute radio frequency channel number}
 \acro{UAV}{unmanned aerial vehicle}
 \acro{UCI}{uplink control information}
 \acro{UDP}{user datagram protocol}
 \acro{UDN}{ultra-dense network}
 \acro{UE}{user equipment}
 \acro{UGS}{unsolicited grant service}
 \acro{UICC}{universal integrated circuit card}
 \acro{UK}{united kingdom}
 \acro{UL}{uplink}
 \acro{UMA}{unlicensed mobile access}
 \acro{UMi}{urban micro}
 \acro{UMTS}{universal mobile telecommunication system}
 \acro{UN}{United Nations}
 \acro{URLLC}{ultra-reliable low-latency communication}
 \acro{US}{upstream}
 \acro{USIM}{universal subscriber identity module}
 \acro{UTD}{theory of diffraction}
 \acro{UTRA}{UMTS terrestrial radio access}
 \acro{UTRAN}{UMTS terrestrial radio access network}
 \acro{UWB}{ultra wide band}
 \acro{VD}{vertical diffraction}
 \acro{VDFP}{vertical dynamic frequency planning}
 \acro{VDSL}{very-high-bit-rate digital subscriber line}
 \acro{VeNB}{virtual eNB}
 \acro{VeNodeB}{virtual eNodeB}
 \acro{VIC}{victim cell}
 \acro{VLR}{visitor location register}
 \acro{VNF}{virtual network function}
 \acro{VoIP}{voice over IP}
 \acro{VoLTE}{voice over LTE}
 \acro{VPLMN}{visited PLMN}
 \acro{VR}{visibility region}
  \acro{VRAN}{virtualized radio access network}
 \acro{WCDMA}{wideband code division multiple access}
 \acro{WEP}{wired equivalent privacy}
 \acro{WG}{working group}
 \acro{WHO}{world health organisation}
 \acro{Wi-Fi}{Wi-Fi}
 \acro{WiMAX}{wireless interoperability for microwave access}
 \acro{WiSE}{wireless system engineering}
 \acro{WLAN}{wireless local area network}
 \acro{WMAN}{wireless metropolitan area network}
 \acro{WNC}{wireless network coding}
 \acro{WRAN}{wireless regional area network}
 \acro{WSEE}{weighted sum of the energy efficiencies}
 \acro{WPEE}{weighted product of the energy efficiencies}
 \acro{WMEE}{weighted minimum of the energy efficiencies}
 \acro{X2}{x2}
 \acro{X2-AP}{x2 application protocol}
 \acro{ZF}{zero forcing}
\end{acronym}

\end{document}